\documentclass[aip,graphicx,reprint,onecolumn,pof, amsmath,amssymb,11pt,floatfix]{revtex4-1}

\draft 

\usepackage{color}
\usepackage{comment}
\usepackage{graphicx}
\usepackage{subcaption}
\usepackage[normalem]{ulem} 

\DeclareMathAlphabet\mathbfcal{OMS}{cmsy}{b}{n}

\graphicspath{ {images/} }

\begin{document}


\title[On the Suppression and Distortion of Non-Equilibrium Fluctuations by Transpiration]{On the Suppression and Distortion of Non-Equilibrium Fluctuations by Transpiration} 



\author{Daniel R. Ladiges, Andrew J. Nonaka, John B. Bell}
\affiliation{Center for Computational Sciences and Engineering,
Lawrence Berkeley National Laboratory}

\author{Alejandro L. Garcia}
\affiliation{Dept. Physics and Astronomy, San Jose State University}

\date{\today} 

\begin{abstract}
A fluid in a non-equilibrium state exhibits long-ranged correlations of its hydrodynamic fluctuations. In this article, we examine the effect of a transpiration interface on these correlations -- specifically, we consider a dilute gas in a domain bisected by the interface. The system is held in a non-equilibrium steady state by using isothermal walls to impose a temperature gradient. The gas is simulated using both direct simulation Monte Carlo (DSMC) and fluctuating hydrodynamics (FHD). For the FHD simulations two models are developed for the interface based on master equation and Langevin approaches. For appropriate simulation parameters, good agreement is observed between DSMC and FHD results with the latter showing a significant advantage in computational speed.
For each approach we quantify the effects of transpiration on long-ranged correlations in the hydrodynamic variables.
\end{abstract}


\maketitle 

\section{Introduction}\label{sec:Intro}

Recent research in the field of fluid dynamics has seen an increased focus on nano and microscale applications, and in particular, the role of thermal fluctuations. These fluctuations are important in areas ranging from microfluidic engineeering\cite{HanggiBrownianMotors} to molecular biology\cite{Kapral2015}. 
Fluctuations also yield information about the transport properties of the fluid itself\cite{berne2013dynamic}, and are important in dynamic processes such as phase transitions, combustion and ignition, nucleation, hydrodynamic instabilities, and a host of Brownian dynamics phenomena.\cite{Lemarchand2,Done3,Vailati1,Lemarchand1,Gallis2015,CasimirPrl2015}

Traditional computational fluid dynamic approaches have focused on solutions to the deterministic Navier-Stokes equations, which neglect thermal fluctuations. To capture this important feature of nanoscale flows particle based methods are typically employed. These techniques operate by replicating, with varying degrees of approximation, the dynamics of the fluid's constituent molecules. Molecular dynamics\cite{Fren1} takes the approach of deterministically simulating the interactions between molecules, often using a relatively complex interaction potential. Although this approach accurately simulates fluids at the nanoscale, its computational cost makes it inappropriate for mesoscale applications where thermal fluctuations are of importance. The direct simulation Monte Carlo\cite{Bird1,Bird8} (DSMC) method greatly simplifies the particle interactions using a stochastic approach, allowing the simulation of much larger systems while still reproducing the correct thermal fluctuations. DSMC solves the Boltzmann equation\cite{Wagn1} so it can be used for flows of arbitrary Knudsen number,\cite{Sone1} whereas solutions of the Navier-Stokes equations are valid only when the Knudsen number is sufficiently small. However this technique applies only to dilute gasses, and is still relatively computationally expensive.

An alternative method for simulating mesoscopic systems is given by fluctuating hydrodynamics (FHD). Following the approach of Landau and Lifshitz \cite{Land2,Zara1}, FHD extends the deterministic Navier-Stokes equations by incorporating stochastic fluxes that lead to fluctuations that are consistent with statistical mechanics. The validity of fluctuating hydrodynamics for non-equilibrium systems has been confirmed experimentally by light scattering\cite{LightFhd1994} and shadowgraph experiments\cite{ShadowFhd2011}. Numerical methods for solving the compressible and incompressible FHD equations are discussed in Refs. \citenum{Bell1, Done1, Balb1}, and the approach has been further extended to multispecies mixtures,\cite{Bell2, Done3, Done4, Bala1, Done5, Nona1, Done6} multiphase flows,\cite{Chau1} reacting flows,\cite{Bhat2} and electrokinetic flows.\cite{Pera6, Pera5}. Hybrid methods combining FHD and DSMC are investigated in Refs. \citenum{Will1,Done2}.

When a fluid is held in a non-equilibrium state, long range correlations in the thermal fluctuations\cite{Zara1} can result in macroscopic effects.\cite{Done5}
These non-equilibrium fluctuations have been studied in many scenarios in which the correlations extend through the bulk of the fluid, however less is known regarding the effect of an interface, such as a porous wall or a thin membrane. This type of interface has previously been studied using molecular dynamics\cite{Broe1} in the context of two separate reservoirs of gas, each held in equilibrium at different temperatures and densities. In this article we consider a dilute gas in a volume bisected by simple porous transpiration interface, with the system driven out of equilibrium by inducing a temperature gradient. The partitioning interface is taken to be a thin adiabatic wall, with holes that are small compared with the mean free path of the gas. This ensures that the pores themselves do not have a significant hydrodynamic effect, yielding a simple system for the study of correlations. 
Particles passing through the interface conserve their energy, but not their momentum, consistent with the particles experiencing randomizing elastic collisions as they pass through the membrane. 

In this article, both DSMC and FHD are used to simulate this system, yielding information on long range correlations and validating the use of FHD for transpiration interfaces. In Section~\ref{sec:Fluid} we begin with a brief review of DSMC, and of the theory underlying fluctuating hydrodynamics including the numerical implementation of the FHD equations.
In Section~\ref{sec:trans} we present several models for transpiration interfaces, using both master equation and Langevin approaches in the FHD context, and the kinetic model applicable to DSMC simulations. In Section~\ref{sec:Results}, we validate these approaches by comparing the long range correlations occurring in both FHD and DSMC simulations, in the presence of interfaces with varying porosity. The role of the interface in distorting and suppressing these correlations is discussed, and a brief comparison of the relative computational performance of the DSMC and FHD simulations is made. Some concluding remarks are given in Section~\ref{sec:Conclusion}.


\section{Fluid Simulation Models}\label{sec:Fluid}

For the study of non-equilibrium fluctuations we use two independent approaches for the bulk fluid: a particle model, Direct Simulation Monte Carlo (DSMC), and a continuum model, fluctuating hydrodynamics (FHD). 
This section summarizes these numerical models; the treatment of the transpiration interface is described in Section~\ref{sec:trans}.

\subsection{Direct Simulation Monte Carlo (DSMC)}\label{sec:DSMC}

Direct Simulation Monte Carlo is a well-known method developed by Graeme Bird for
computing gas dynamics from the perspective of kinetic theory.\cite{Bird1} 
As in molecular dynamics, the state of the system in DSMC is given by the positions and velocities of particles. At each time step the particles are first
independently moved ballistically, imposing boundary conditions if they strike a surface. Generally, isothermal boundary conditions are implemented using a diffuse surface interaction while adiabatic boundaries are simulated using specular reflection; surfaces with partial accommodation can also be simulated. Collisions are evaluated by a stochastic process, selecting the post-collision angles from their kinetic theory distributions while conserving momentum and energy. We employ the hard sphere collision operator in the present work but many other collisions models are available in DSMC.
Pedagogical explanations of DSMC are contained in Refs.~\citenum{Alex2}, \citenum{Garc2}, and \citenum{Boyd1} with a complete description given in Ref.~\citenum{Bird8}.

Although DSMC is a stochastic algorithm the “Monte Carlo” elements of the method have nothing to do with the statistical variation of the hydrodynamic variables; the same fluctuations also occur in deterministic methods, such as molecular dynamics. Under a relatively weak set of conditions (e.g., detailed balance) discrete event processes reproduce the correct physical fluctuations at hydrodynamic scales.\cite{Keiz1,Keiz2}
When each DSMC particle represents one molecule in the gas, the magnitude of the hydrodynamic fluctuations is in agreement with equilibrium statistical mechanics.\cite{Hadj6}
For both equilibrium and non-equilibrium systems DSMC yields the physical spectra of spontaneous fluctuations as predicted by theory\cite{Mans1,Garc4} and observed in experiments\cite{Bruno1,Bruno2}.
In this paper the numerical measurements from DSMC simulations are used to validate the fluctuating hydrodynamic simulation results.

\subsection{Fluctuating Hydrodynamics (FHD)}\label{sec:FHD}

The equations of fluctuating hydrodynamics are obtained by adding white noise to each dissipative flux in the deterministic Navier-Stokes equations,\cite{Giov1} to represent thermal fluctuations. We consider the compressible fluctuating hydrodynamic equations for a pure monatomic ideal gas; generalizations are found in Refs. \citenum{Bell2, Done3, Done4, Bala1, Done5, Nona1, Done6,Chau1,Bhat2,Pera6}.
The continuity, momentum and energy conservation equations for FHD are
\begin{equation}
\frac{\partial }{\partial t} \left( \rho \right)  = - {\bf \nabla} \cdot \left( \rho {\bf u} \right),
\label{eqn:cont}
\end{equation}
\begin{equation}
\frac{\partial }{\partial t} \left( \rho {\bf u} \right) =  - {\bf \nabla} \cdot \left( \rho {\bf u \otimes u } \right) - {\bf \nabla} P - {\bf \nabla} \cdot \left[ {\bf \Pi} + {\bf \widetilde{\Pi}} \right],
\label{eqn:mom}
\end{equation}
\begin{eqnarray}
\frac{\partial }{\partial t} \left( \rho E \right)  = - {\bf \nabla} \cdot \left( \rho {\bf u} E + p {\bf u}  \right) - {\bf \nabla} \cdot \left[ {\bf Q} + {\bf \widetilde{Q}} \right] -
{\bf \nabla} \cdot \left( {\bf \left[ \Pi + \widetilde{\Pi} \right] \cdot  u} \right).
\label{eqn:energy}
\end{eqnarray}
Here, $\rho$ is the mass density of the gas, $\bf{u}$ is the velocity, $P$ is the pressure, $T$ is temperature, and $E = c_V T + \frac{1}{2}|{\bf u}|^2$ is the total specific energy. For a monatomic ideal gas the specific heat at constant volume is $c_V = \frac32 k_B /m$, where $k_B$ is Boltzmann's constant and $m$ is the molecular mass; the equation of state is $P = n k_B T$ where $n=\rho/m$ is the number density.
From Eqs.~(\ref{eqn:cont})-(\ref{eqn:energy}) we define the fluxes in the mass, momentum, and total energy as $J_M$, $J_{\mathbfcal{P}}$, and $J_{\mathcal{E}}$, respectively.

In the momentum and energy equations (Eqs.~(\ref{eqn:mom}) and ~(\ref{eqn:energy})), the deterministic viscous stress tensor ${\bf \Pi}$ and deterministic heat flux vector ${\bf Q}$ are given by 
\begin{equation}
\Pi_{ij} = -\eta \left( \frac{\partial u_i}{\partial x_j} + \frac{\partial u_j}{\partial x_i}  \right) + \delta_{ij} \left( \frac{2}{3} \eta {\bf \nabla} \cdot {\bf u} \right),
\qquad\mathrm{and}\qquad
{\bf Q} =  -  \kappa {\bf \nabla} T,
\label{eqn:Q}
\end{equation}
where $\delta_{ij}$ is the Kronecker delta, $\eta(T)$ is the shear viscosity, and $\kappa(T)$ is the thermal conductivity; consistent with a monatomic gas, we take the bulk viscosity to be zero.

The terms that appear with a tilde in Eqs.~(\ref{eqn:mom}) and (\ref{eqn:energy}) denote stochastic contributions from Gaussian random fields. The stochastic viscous tensor, $\mathbf{\widetilde{\Pi}}$, stochastic heat flux tensor, $\mathbf{\widetilde{Q}}$, 
have zero mean and covariances~\cite{Land2,Zara1}
\begin{equation}
\langle \widetilde{{\Pi}}_{ij}(\mathbf{r},t) \; , \widetilde{{\Pi}}_{kl}(\mathbf{r}',t') \rangle
= 2 k_B \eta T \left [ ( \delta_{ik}\delta_{jl}+ \delta_{il}\delta_{jk} )
- {\textstyle \frac13} \delta_{ij}\delta_{kl} \right ]
 \delta(t-t') \delta( \mathbf{r} - \mathbf{r}' ),\label{eqn:pistoch}
\end{equation}
and
\begin{equation}
\langle \widetilde{{Q}}_{i}(\mathbf{r},t) \; , \widetilde{{Q}}_{j}(\mathbf{r}',t') \rangle = 2 k_B \kappa T^2  \delta_{ij} \delta(t-t') \delta( \mathbf{r} - \mathbf{r}' ).
\label{eqn:qstoch}
\end{equation}
where $\delta(..)$ is the Dirac delta function.

The FHD Eqs.~(\ref{eqn:cont})-(\ref{eqn:energy}) above, are simulated numerically in one dimension using a finite volume method with  third-order Runge-Kutta time stepping. A complete description of this approach is given in Sec VI of Ref.~\citenum{Done1}.

\section{Transpiration Interface Models}\label{sec:trans}

As discussed in Section \ref{sec:Intro}, we consider an adiabatic, stationary interface that is porous to the gas molecules (e.g., a specular wall with small holes). Particles encountering the interface are allowed to pass with an effusion probability, $f$. This probability can be viewed as the fraction of the wall area comprised of holes. The conditions in the gas are chosen such that a particle that crosses the interface experiences a large number of inter-molecular collisions before crossing again -- this is called the effusion condition\cite{Bird1,Liep1}. 

\subsection{DSMC interface model}\label{sec:dsmcinterface}
The implementation of the transpiration interface is straightforward in DSMC.  Particles reaching the interface are transmitted with probability $f$ and specularly reflected with probability $1-f$. 
The transmitted particles are assumed to interact with the adiabatic porous interface such that the particle's energy, $\epsilon$, is conserved but not its momentum. Therefore the velocity of a transmitted particle, $\mathbf{v}$, is reset by randomly selecting it from a hemisphere of radius $|\mathbf{v}| = \sqrt{2 \epsilon/m}$. 

\subsection{FHD interface models}\label{sec:fhdinterface}

The specular condition for the surface of the interface, discussed above, is modeled hydrodynamically as a full slip adiabatic boundary condition. We present two methods
for modeling the effusion of mass and energy through the interface, based on master equation and Langevin approaches; as discussed above, the effusive momentum flux is set to zero. We first discuss the kinetic theory underlying both these techniques.

\subsubsection{Kinetic theory}\label{sec:kinetictheory}

For a gas at equilibrium, the time between particle crossings through a transpiration interface of area $A$ is exponentially distributed, with the mean time between crossings given by\cite{Bird1}
\begin{equation}
\tau_\mathrm{e} = \frac{m}{\rho f A}\sqrt{\frac{2\pi m}{k_B T}}.\label{eqn:meantime}
\end{equation}
By comparison, the mean time between collisions for a hard sphere particle\cite{Bird1} is
\begin{equation}
\tau_\mathrm{c} = \frac{m}{4 \sqrt{\pi} \rho d^2}\sqrt{\frac{2\pi m}{k_B T}},
\end{equation}
where $d$ is the collisional diameter of the particle. The effusion condition described above is $n A \lambda_\mathrm{c} \tau_\mathrm{e} \gg \tau_\mathrm{c}$ where $\lambda_\mathrm{c}$ is the mean free path, which is satisfied when $f \ll 1$. 
The probability distribution for the kinetic energy, $\epsilon$, of particles passing through the interface is,
\begin{equation}
P_\epsilon (\epsilon) = \frac{\epsilon}{k_B^2 T^2} ~e^{\displaystyle  -\epsilon/(k_B T)}; \label{eqn:energydist}
\end{equation}
this is a gamma distribution, $\Gamma(2,k_B T)$.\cite{Gent1} Eqs.~(\ref{eqn:meantime}) and (\ref{eqn:energydist}) form the basis of the stochastic simulation algorithm given in Section~\ref{sec:masterimplement}.

From the above we find that the probability per unit time of observing a particle of energy $\epsilon$ crossing the interface is
\begin{equation}
W_\rightarrow(\epsilon) = \frac{\displaystyle f A\,  n_L }{\sqrt{2\pi m k_B T_L}}\frac{\epsilon}{k_B T_L}~e^{\displaystyle  -\epsilon/(k_B T_L)},
\end{equation}
\begin{equation}
W_\leftarrow(\epsilon) = \frac{\displaystyle f A\,  n_R }{\sqrt{2\pi m k_B T_R}}\frac{\epsilon}{k_B T_R}~e^{\displaystyle  -\epsilon/(k_B T_R)},
\end{equation}
where the arrows indicate the direction of the crossing, and the subscripts $L$ and $R$ refer to the gas on the left and right hand sides of the interface. This leads to the master equation for $P(M,\mathcal{E})$, i.e., the probability density of the state on the right of the interface with mass $M$ and total energy $\mathcal{E}$,
\begin{align}
\frac{d}{d t}P(M, \mathcal{E})=\int_0^\infty P(M-m,\mathcal{E}-\epsilon) &W_\rightarrow(\epsilon) +  P(M+m,\mathcal{E}+\epsilon) W_\leftarrow(\epsilon) d\epsilon\nonumber\\
-&\int_0^\infty P(M,\mathcal{E}) W_\rightarrow(\epsilon) +  P(M,\mathcal{E}) W_\leftarrow(\epsilon) d\epsilon.
\end{align}


Following from the above,~\cite{Broe1} the average one-way flux of mass passing from left to right through the interface is,
\begin{equation}
\langle J_{M}\rangle_\rightarrow = \rho_L f A~\sqrt{\frac{k_B T_L}{2\pi m}},
\end{equation}
with the arrow indicating the direction of flux. The total average mass flux is given by
\begin{equation}
\langle J_{M}\rangle= \langle J_{M}\rangle_\rightarrow -\langle J_{M}\rangle_\leftarrow
=  \frac{\displaystyle f A\, k_B^{1/2}}{\displaystyle\sqrt{2 \pi m  }}\left(\rho_L T^{1/2}_L- \rho_R T^{1/2}_R  \right).\label{eqn:meanmass}
\end{equation}
 The average energy of a particle crossing the interface is $2 k_B T$, so the average total energy flux is
\begin{align}
\langle J_{\mathcal{E}}\rangle= \frac{2 k_B }{m}~
\left( T_L \langle J_{M}\rangle_\rightarrow- T_R \langle J_{M}\rangle_\leftarrow\right)=\frac{\displaystyle 2 f A\, k_B^{3/2}}{\displaystyle m \sqrt{2 \pi m  }}\left(\rho_L T^{3/2}_L - \rho_R T^{3/2}_R  \right).\label{eqn:meanenergy}
\end{align}
As discussed above, we assume that there is no momentum flux across the interface so $J_\mathbfcal{P}=0$. Since the non-equilibrium conditions considered in this article result in zero average net momentum transport, this difference in treatment between DSMC and FHD appears to have a negligible impact on the long-ranged correlations of interest (see Section~\ref{sec:CompareFhdDsmc}).

The variances and covariance of the mass and energy fluxes are
\begin{equation}
\langle J_{\delta M\delta M} \rangle = \frac{\displaystyle m f A\, k_B^{1/2}}{\displaystyle\sqrt{2 \pi m  }}\left(\rho_R T^{1/2}_R+\rho_L T^{1/2}_L  \right)\label{eqn:varmass},
\end{equation}
\begin{equation}
\langle J_{\delta \mathcal{E}\delta \mathcal{E}} \rangle = \frac{\displaystyle 6 f A\, k_B^{5/2}}{\displaystyle m\sqrt{2 \pi m  }}\left(\rho_R T^{5/2}_R+\rho_L T^{5/2}_L  \right)\label{eqn:varenergy},
\end{equation}
\begin{equation}
\langle J_{\delta\mathcal{E} \delta M} \rangle = \frac{\displaystyle 2 f A\, k_B^{3/2}}{\displaystyle\sqrt{2 \pi m }}\left(\rho_R T^{3/2}_R+\rho_L T^{3/2}_L  \right)\label{eqn:covarmassenergy}.
\end{equation}
The resulting Langevin equations are given by
\begin{equation}
\frac{d }{dt}M_R=\, \langle J_{M} \rangle + \widetilde{J}_M, \nonumber\\
\end{equation}
\begin{equation}
\frac{d }{dt}{\mathcal{E}_R}=\, \langle J_{\mathcal{E}} \rangle+\widetilde{J}_\mathcal{E},\nonumber\\
\end{equation}
\begin{equation}
\frac{d }{dt}{M_L}=\, -\frac{d }{dt}{ M_R},
\quad 
\frac{d }{dt}{\mathcal{E}_L}=\, -\frac{d }{dt}{\mathcal{E}_R},
\label{eqn:langevin}
\end{equation}
where $\widetilde{J}_M$ and $\widetilde{J}_\mathcal{E}$ are Gaussian noise terms, with
\begin{equation}
\langle \widetilde{J}_M(t)\widetilde{J}_M(t')\rangle=\langle J_{\delta M\delta M} \rangle\delta(t-t'),\nonumber
\end{equation}\begin{equation}
\langle \widetilde{J}_\mathcal{E}(t)\widetilde{J}_\mathcal{E}(t')\rangle=\langle J_{\delta \mathcal{E}\delta \mathcal{E}} \rangle\delta(t-t'),\nonumber
\end{equation}\begin{equation}
\langle \widetilde{J}_\mathcal{E}(t)\widetilde{J}_M(t')\rangle=\langle J_{\delta \mathcal{E}\delta M} \rangle\delta(t-t').
\end{equation}
These equations form the basis of the Langevin method discussed in Section~\ref{sec:langevinimplement}.


\subsubsection{Master equation implementation}\label{sec:masterimplement}
Following from the theory presented above, we define a stochastic simulation algorithm (SSA)\cite{Gill1} implementation of the interface. This operates by repeatedly generating particle crossing times from an exponential distribution, until the cumulative sum of these crossing times exceeds a single FHD time step, $\Delta t$. The energy of each crossing particle is selected from the appropriate gamma distribution, and a tally of the total mass and energy transfer is used to update the hydrodynamic cells on each side of the interface. 

The implementation of the master equation model over a single time step is as follows:
\begin{itemize}
\item Take $\rho_L$ and $\rho_R$ to be the density at the cell centers to the left and to the right of the interface. Similarly, the temperature and specific energy in said cells is indicated by $T_L$, $T_R$, $E_L$, and $E_R$.
\item From Eq.~(\ref{eqn:meantime}) the rate of particles crossing from each direction is $r_L = 1/\tau_\mathrm{e}(\rho_L,T_L)$ and $r_R = 1/\tau_\mathrm{e}(\rho_R,T_R)$.
\item Initialize the accumulation of mass and energy flux, $F_M$ and $F_\mathcal{E}$, across the interface to zero.
\item Set the interface time counter $t_\mathrm{ssa} = 0$. Loop while $t_\mathrm{ssa} < \Delta t$:
\begin{itemize}
\item Given an exponential distribution of time
between particle crossings, the time until the next particle crossing is given by
\begin{equation}
\theta = -\frac{\displaystyle \log(\mathcal{R}_1)}{\displaystyle r_L + r_R},
\end{equation}
where $\mathcal{R}_i \in (0,1]$ indicates an independent uniform random number.
\item Increment the interface time counter, $t_\mathrm{ssa} \mathrel{+}= \theta$. If $t_\mathrm{ssa} \geq \Delta t$ then break out of the loop.
\item Select the direction for the particle crossing the interface. If $r_L/(r_L + r_R) > \mathcal{R}_2$, then the particle crosses from left to right, otherwise it crosses from right to left.
\item Following from Eq.~(\ref{eqn:energydist}), the energy of the effusing particle is gamma distributed and generated by\cite{Gent1}
\begin{equation}
\epsilon = -k_B T_* \log(\mathcal{R}_3\mathcal{R}_4),
\end{equation}
where $T_* = T_L$ if the particle is traveling left to right, or $T_* = T_R$ otherwise.
\item If the particle is traveling left to right, $F_M \mathrel{+}= m$ and $F_{\mathcal{E}} \mathrel{+}= \epsilon$. Otherwise $F_M \mathrel{-}= m$ and $F_\mathcal{E} \mathrel{-}= \epsilon$.
\end{itemize}
\item Finally, the left cell is updated as $\rho_L \mathrel{-}= F_M /\Delta V$ and $\rho_LE_L \mathrel{-}= F_\mathcal{E} /\Delta V$, and the right cell is updated as $\rho_R \mathrel{+}= F_M /\Delta V$ and $\rho_R E_R \mathrel{+}= F_\mathcal{E} /\Delta V$ with $\Delta V$ being the cell volume.
\end{itemize}

Note that the interface is modeled as a slip wall since the flux of momentum crossing the interface is zero.

\subsubsection{Langevin equation implementation}\label{sec:langevinimplement}

In contrast to the SSA approach, which considers discrete particle crossing events, the Langevin algorithm generates the total mass and energy transfer based on the expected values. A Gaussian approximation is used for the fluctuations in these quantities -- this is similar in principle to the stochastic fluxes generated between cells in the FHD algorithm. 
The implementation of the Langevin model over a single time step is as follows:
\begin{itemize}
\item Take $\rho_L$ and $\rho_R$ to be the density at the cell centers to the left and to the right of the interface. Similarly, the temperature and specific energy in said cells is indicated by $T_L$, $T_R$, $E_L$, and $E_R$.
\item Calculate the mean mass and energy fluxes given by Eqs.~(\ref{eqn:meanmass}) and (\ref{eqn:meanenergy}), and the variances and covariance given by  Eqs.~(\ref{eqn:varmass})-(\ref{eqn:covarmassenergy}).
\item Calculate the correlation coefficient, given by
\begin{equation}
r=\frac{\langle J_{\delta \mathcal{E}\delta M} \rangle}{\sqrt{\langle  J_{\delta\mathcal{E}\delta \mathcal{E}} \rangle\langle  J_{\delta M\delta M} \rangle}}.
\end{equation}
\item  Generate correlated random variables,
\begin{equation}
\mathcal{G}_3 = \mathcal{G}_1 r + \sqrt{1-r^2} \mathcal{G}_2
\end{equation}
where $\mathcal{G}_1$ and $\mathcal{G}_2$ are independent Gaussian random numbers of mean zero and variance one.
\item Following from Eq.~(\ref{eqn:langevin}), the mass and energy flux accumulations are given by
\begin{equation}
F_M = \Delta t \langle  J_{M} \rangle + \sqrt{\Delta t\langle  J_{\delta M\delta M} \rangle}~\mathcal{G}_1,
\end{equation}
\begin{equation}
F_\mathcal{E} = \Delta t \langle  J_{\mathcal{E}} \rangle + \sqrt{\Delta t\langle  J_{\delta\mathcal{E}\delta \mathcal{E}} \rangle}~\mathcal{G}_3.
\end{equation}
\item Finally, the left cell is updated as $\rho_L \mathrel{-}= F_M /\Delta V$ and $\rho_LE_L \mathrel{-}= F_\mathcal{E} /\Delta V$, and the right cell is updated as $\rho_R \mathrel{+}= F_M /\Delta V$ and $\rho_R E_R \mathrel{+}= F_\mathcal{E} /\Delta V$ with $\Delta V$ being the cell volume.
\end{itemize}

Again, note that the interface is modeled as a slip wall since the flux of momentum crossing the interface is zero.

\section{Simulation Results} \label{sec:Results}
In this section, we begin by validating the use of FHD for transpiration by comparison of the FHD and DSMC simulations. For the majority of this section, the master equation method of Section~\ref{sec:masterimplement} is used to implement the interface in the FHD simulations; several comparisons are shown in Section~\ref{sec:CompareFhdDsmc} to validate the Langevin method of Section~\ref{sec:langevinimplement}. The effect of transpiration is demonstrated with several different effusion probabilities, $f$, by examining the long range spatial correlations in the fluid. This is followed by an analysis of the relative computational efficiency of the DSMC and FHD simulations.

\begin{figure}
\centering
\includegraphics[width=0.9\linewidth]{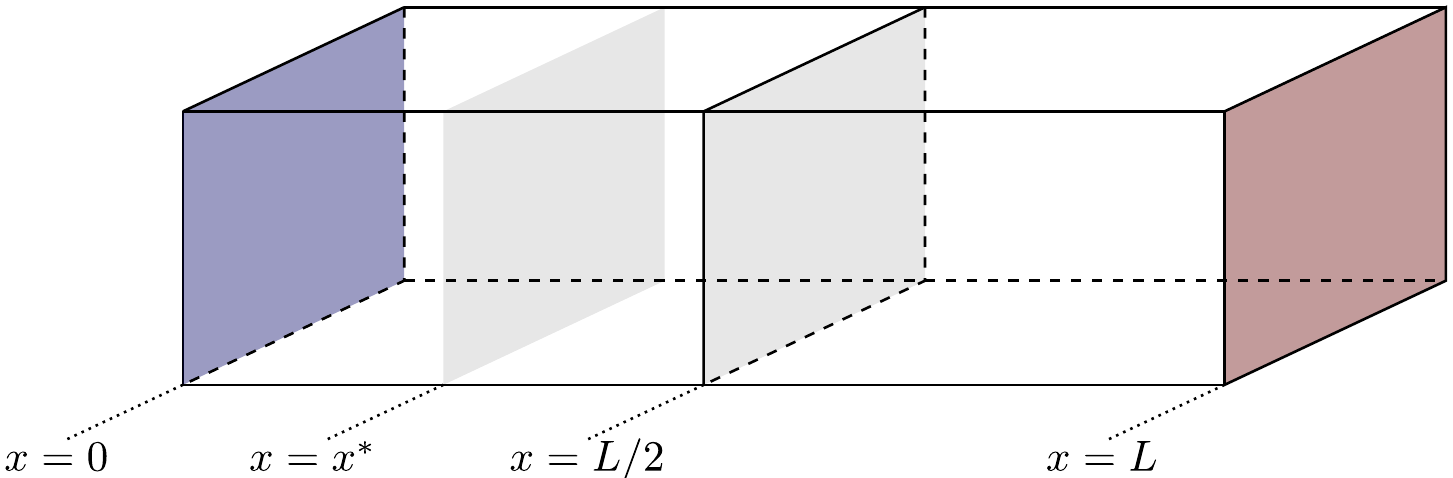}
\caption{Illustration of the simulation geometry. Walls at $x=0$, $L$ held at fixed temperatures. Transpiration interface (membrane) located at $x=L/2$ and correlations of fluctuations are measured relative to position $x^* \approx L/4$. }\label{fig:diag}
\end{figure}

We consider a quasi-one dimensional domain with a gradient induced in the gas by setting different temperatures at $x=0$ and $x=L$ (see Fig.~\ref{fig:diag}). These temperatures are set in  the DSMC simulations using a fully diffuse boundary condition, and in the FHD simulations with an isothermal no-slip boundary condition. Rather than using a temperature slip boundary condition, the Dirichlet boundary condition for temperature is set to match the mean extrapolated temperatures from the DSMC simulations -- this is discussed further below. For the DSMC simulations the $y$ and $z$ boundaries are taken to be periodic; as discussed above, the FHD simulations are performed in 1D. The transpiration interface is at  $x=L/2$, the center of the domain. Simulation parameters are given in Table~\ref{table:parameters}.


\begin{table}[ht]
\begin{tabular}{l  l c  l l}
&\vspace{-4mm}&&&\\
\cline{1-2}\cline{4-5}
\vspace{-2mm}&\\
Molecular diameter \hspace{20mm} & $3.66\times 10^{-8}$ 
& $\,\,$\hspace{15mm}&
System length ($x$ dir) \hspace{13mm}& $2.5\times 10^{-4}$ \\
Molecular mass & $6.63\times 10^{-23}$ 
&&
Interface $x$ location & $1.25\times 10^{-4}$ \\
Reference temperature & $273$ 
&&
Cell size $\Delta x$ & $3.125\times 10^{-6}$\\
Sound speed & $30,781$ 
&&
Time step $\Delta t$ & $1.0\times 10^{-12}$\\
Specific heat $c_v$ & $3.12\times 10^{6}$ 
&&
Number of cells ($x$ dir) & 80\\
Reference mean free path  & $6.26\times 10^{-6}$ 
&&
System height, width ($y, z$ dir)& $1.25\times 10^{-6}$  \\
Reference mean free time& $1.64\times 10^{-10}$
&&
Average particles per cell & $131$\\
\vspace{-3mm}&\\
\cline{1-2}\cline{4-5}
\vspace{-4mm}&\\

\end{tabular}
\caption{Parameters used when comparing FHD and DSMC simulations, corresponding to a dilute argon gas, in cgs units. For DSMC simulations the molecular properties are applied directly, for FHD simulations they are used to calculate the viscosity and thermal conductivity (see Ref.~\citenum{Bird1}) 
}
\label{table:parameters}
\end{table}

\subsection{Temperature profiles and effusion condition}
\label{sec:EffusionCondition}

\begin{figure}[t]
\centering
\includegraphics[width=0.95\linewidth]{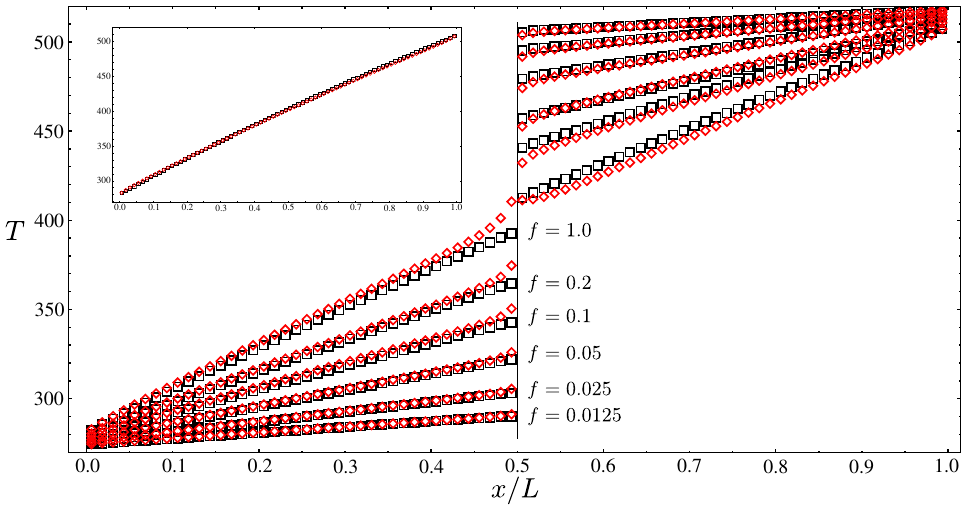}
\caption{Mean temperature versus position for various values of the effusion probability, $f$. The red diamonds indicate the DSMC solution, and the black squares show the FHD result using the master equation method for the interface. The inset plot shows the case when there is no transpiration interface in the system. Because the interface transmits no momentum, this is distinct from the $f=1.0$ case.} \label{fig:fullgrad}
\end{figure}

In Fig.~\ref{fig:fullgrad}, the temperature profiles are compared between the DSMC and FHD simulations (using the master equation algorithm of Section~\ref{sec:masterimplement}). Comparisons are shown for a range of $f$ values, and in the absence of a membrane. In all cases the temperature boundary conditions for the DSMC simulations are set to 273K and 519K on the left and right hand boundaries, respectively, giving a large gradient of $7604\, \mathrm{K}/\mathrm{cm}$. Due to this large gradient, the gas at the boundary is unable to completely accommodate to the wall temperature, resulting in a significant Knudsen layer\cite{Sone1} and `temperature slip'. This feature is correctly captured by the DSMC simulations, but must be accounted for in the FHD simulations with a modified boundary condition.  As mentioned above, the FHD boundary conditions have been extrapolated from these new temperatures; see Table~\ref{table:temperatures}. 

In all cases the DSMC and FHD simulations show good agreement aside from a small discrepancy near the interface, which becomes prominent at $f \gtrsim 0.1$. As discussed in sections~\ref{sec:Intro} and \ref{sec:trans}, the transpiration models introduced for the FHD simulations are valid only when the effusion condition holds. Similar to the temperature boundary condition discussed above, this is highlighted by the appearance of a Knudsen layer at the surface of the interface, with the relevant Knudsen number being proportional to the effusion probability $f$. For $f \gtrsim 0.1$, this causes the solution near the interface to be poorly approximated by the FHD simulations. This occurs because in this regime there is a substantial chance of a particle crossing the interface in one direction and being reflected back in the other direction after only a small number of collisions, preventing the regions close to the interface from maintaining local equilibrium.  This effect is accurately captured by DSMC, which does not rely on a local equilibrium assumption, but not by the FHD simulations. 


\begin{table}[ht]
\begin{tabular}{l|l l l l}
$f$  && $T(0)$ && $T(L)$\\
\hline
no membrane \hspace{15mm}&& 280.8 && 508.5\\
0.1 &\hspace{5mm}& 277.6 &\hspace{5mm}& 509.9 \\
0.05 &&  276.2 && 515.4\\
0.025 && 275.0 && 516.9\\ 
0.0125 &&  274.1 && 517.9\\
0.00625 &&  273.6 && 518.4 \\ 
\end{tabular}
\caption{Boundary temperatures extrapolated from DSMC results, for use in FHD simulations. The DSMC wall temperatures were 273~K and 519~K.
}
\label{table:temperatures}
\end{table}

\subsection{Measurement of hydrodynamic fluctuations}
The fluctuations of the hydrodynamic variables are measured by accumulating statistical samples of the conserved variables. For example, in DSMC the mass and momentum densities in the cell centered at $x$ at time $t$ are,
\begin{equation}
\rho(x,t) = \frac{m}{\Delta V} N(x,t)
\qquad\mathrm{and}\qquad
\mathbf{p}(x,t) = \frac{m}{\Delta V} \sum_i^N \mathbf{v}_i,
\end{equation}
where $N$ is the number of DSMC particles in the cell, and $\mathbf{v}_i$ is the velocity of particle $i$. In FHD these conserved densities, $\rho$ and $\mathbf{p} = \rho \mathbf{u}$, 
are computed directly (see Eqns.~\ref{eqn:cont}-\ref{eqn:energy}). Means and second moments are computed from these statistical samples, for example,
\begin{equation}
\langle \mathbf{p}(x^*) \rho(x) \rangle 
= \frac{1}{S} \sum_{t}^S \mathbf{p}(x^*,t) \rho(x,t)
\end{equation}
and
\begin{equation}
\langle \delta \mathbf{p}^* \delta \rho \rangle =
\langle \delta \mathbf{p}(x^*) \delta \rho(x) \rangle =
\langle \mathbf{p}(x^*) \rho(x) \rangle
- \langle \mathbf{p}(x^*) \rangle \langle \rho(x) \rangle,
\end{equation}
where $S$ is the number of time samples. Correlations of hydrodynamic fluctuations are obtained from these conserved variables, e.g.,
\begin{equation}
\langle \delta \mathbf{u}^* \delta \rho \rangle = 
\frac{1}{\langle \rho^* \rangle} \Big( 
\langle \delta \mathbf{p}^* \delta \rho \rangle
- \langle \mathbf{u}^* \rangle \langle \delta \rho^* \delta \rho \rangle
\Big).
\end{equation}
The calculation of temperature fluctuations is similar; see Ref.~\citenum{Garc3} for details. For all the results shown in this article the simulation was run until a statistical steady state was reached, recording of the quantities of interest was then begun. The value $S$ is given in the caption for each result. In practice each statistical sample was constructed from an ensemble of shorter simulations to give a total of $S$ time samples.

\subsection{Hydrodynamic fluctuations in DSMC and FHD}
\label{sec:CompareFhdDsmc}

\begin{figure}
\centering

\begin{subfigure}[t]{.45\textwidth}
\centering
\includegraphics[width=\linewidth]{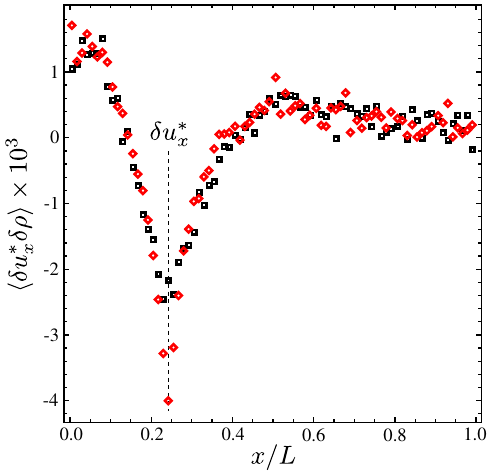}
        \caption{}\label{fignomembrane:velrho}
\end{subfigure}
\begin{subfigure}[t]{.45\textwidth}
\centering
\includegraphics[width=\linewidth]{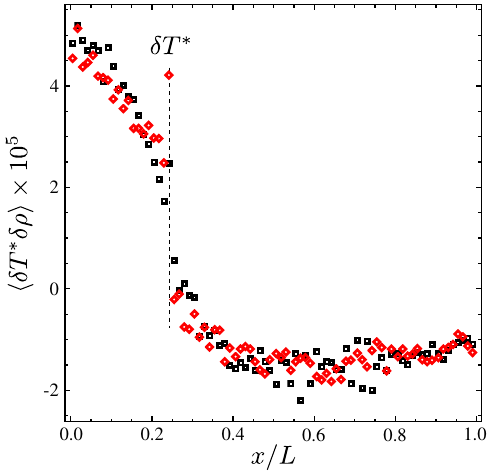}
\caption{}\label{fignomembrane:temprho}
\end{subfigure}

\medskip

\begin{subfigure}[t]{.45\textwidth}
{\centering
\vspace{0pt}
\includegraphics[width=\linewidth]{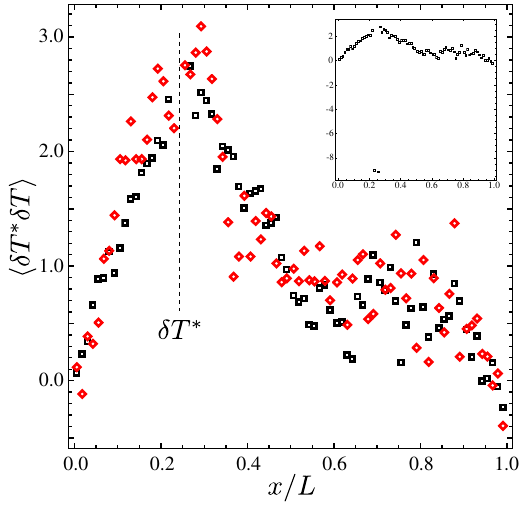}}
\caption{}\label{fignomembrane:temptemp}
\end{subfigure}
\begin{minipage}[t]{.05\textwidth}
\hspace{3mm}
\end{minipage}
\begin{minipage}[t]{.4\textwidth}

\caption{Spatial correlations in the absence of a interface. DSMC results are indicated by the red diamonds, FHD results are given by the black squares. In (\subref{fignomembrane:velrho}), the correlation is shown between the velocity fluctuations at location $x^*$ and the density fluctuations at every other cell. The correlation between temperature fluctuations at $x^*$ and the density fluctuations in other cells is shown in (\subref{fignomembrane:temprho}), and the temperature-temperature fluctuations are shown in (\subref{fignomembrane:temptemp}). Each plot was generated using $S=4\times 10^{8}$ samples. Good agreement is observed in all cases, confirming the
findings of Ref.~\citenum{Bell1}.
}\label{fig:compnomembrane}
\end{minipage}

\end{figure}

In Fig.~\ref{fig:compnomembrane} long range spatial correlations, in the absence of an interface, are compared between DSMC and FHD simulations. Three correlations are shown, $\langle \delta u_x^* \delta \rho \rangle$, $\langle \delta T^* \delta \rho \rangle$, and $\langle \delta T^* \delta T \rangle$, in (a), (b), and (c), respectively. In each case the variable with the asterisk is measured at the fixed location $x^* = L/4-\Delta x/2$, i.e., in the cell near the center of the left hand side of the domain; see Fig.~\ref{fig:diag}. Note that in the absence of a temperature gradient (i.e., at thermodynamic equilibrium) $\langle \delta u_x^* \delta \rho \rangle = \langle \delta T^* \delta \rho \rangle = 0$ and $\langle \delta T^* \delta T \rangle = 0$ for $x \neq x^*$.

We see good agreement between the FHD and DSMC results, observing that the FHD simulations have difficulty capturing the sharp peak in $\langle \delta u_x^* \delta \rho \rangle$; this confirms the observations of Ref.~\citenum{Bell1}. In (c), the expected Kroenecker delta peak at $x=x^*$ has been removed, from both the FHD and DSMC results, for ease of viewing. The FHD simulations again have difficulty in this case, and exhibit significant undershoot in the two adjacent cells. This is a well known effect,\cite{Garcia1987} and these data points have also been removed from the FHD results for ease of viewing -- they are shown on the inset plot.

\begin{figure}
\centering

\begin{subfigure}[t]{.45\textwidth}
\centering
\includegraphics[width=\linewidth]{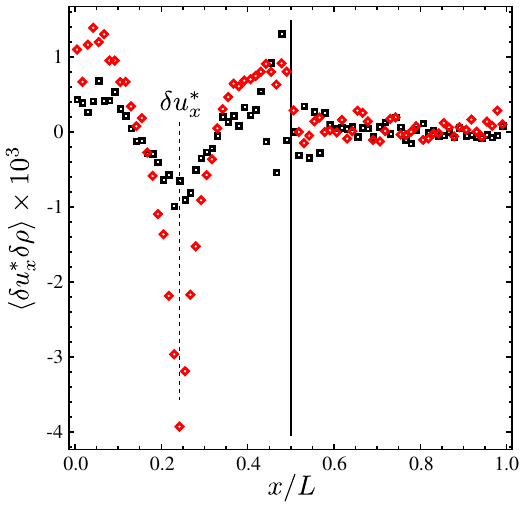}
        \caption{}\label{fig01:velrho}
\end{subfigure}
\begin{subfigure}[t]{.45\textwidth}
\centering
\includegraphics[width=\linewidth]{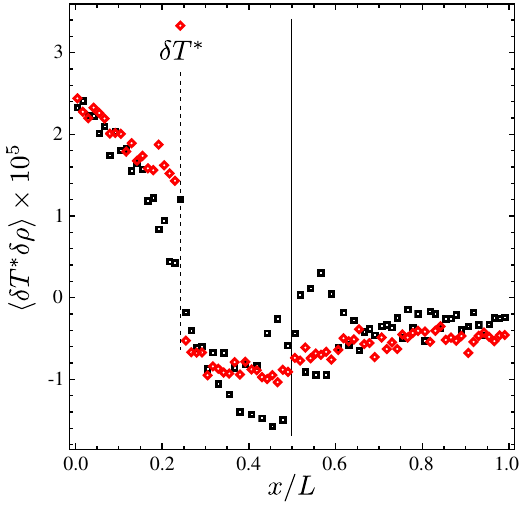}
\caption{}\label{fig01:temprho}
\end{subfigure}

\medskip

\begin{subfigure}[t]{.45\textwidth}
\centering
\vspace{0pt}
\includegraphics[width=\linewidth]{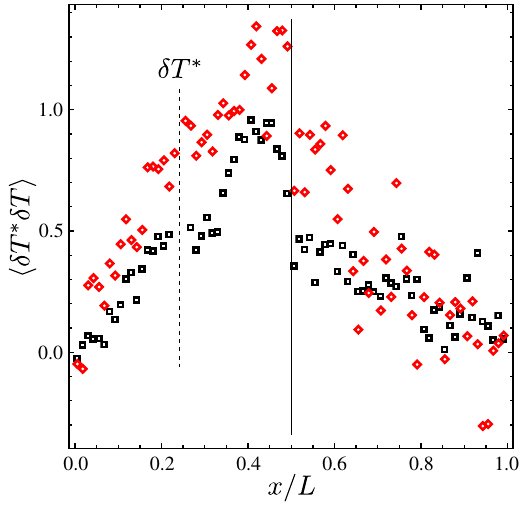}
\caption{}\label{fig01:temptemp}
\end{subfigure}
\begin{minipage}[t]{.05\textwidth}
\hspace{3mm}
\end{minipage}
\begin{minipage}[t]{.4\textwidth}

\caption{Spatial correlations in the presence of a interface with $f=0.1$. DSMC results are indicated by the red diamonds, FHD results using a master equation for the interface are given by the black squares. In (\subref{fig01:velrho}), the correlation is shown between the velocity fluctuations at location $x^*$ and the density fluctuations at every other cell. The correlation between temperature fluctuations at $x^*$ and the density fluctuations in other cells is shown in (\subref{fig01:temprho}), and the temperature-temperature fluctuations are shown in (\subref{fig01:temptemp}). Each plot was generated using $S=1.6\times 10^{9}$ samples. Approximate agreement is seen in each case, with some differences in magnitude visible -- this is due to the violation of the effusion condition.}\label{fig:comp01}
\end{minipage}

\end{figure}

\begin{figure}
\centering

\begin{subfigure}[t]{.45\textwidth}
\centering
\includegraphics[width=\linewidth]{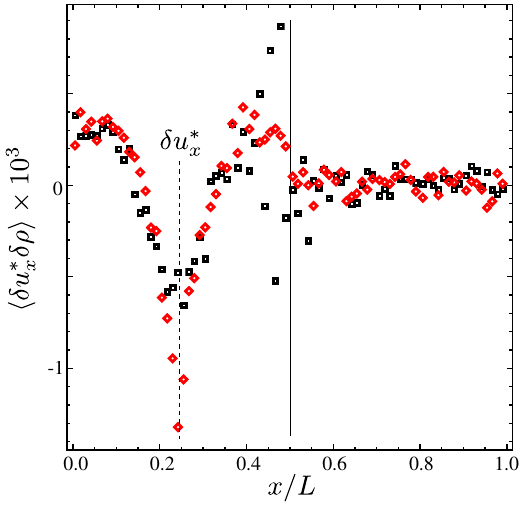}
        \caption{}\label{fig005:velrho}
\end{subfigure}
\begin{subfigure}[t]{.45\textwidth}
\centering
\includegraphics[width=\linewidth]{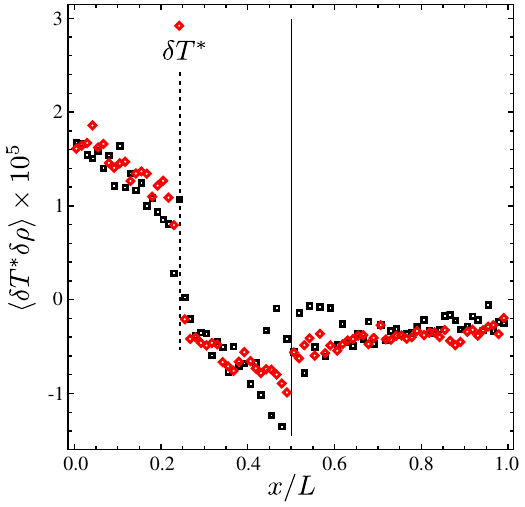}
\caption{}\label{fig005:temprho}
\end{subfigure}

\medskip

\begin{subfigure}[t]{.45\textwidth}
\centering
\vspace{0pt}
\includegraphics[width=\linewidth]{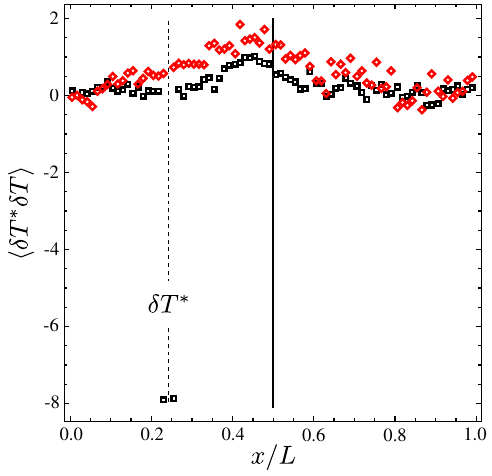}
\caption{}\label{fig005:temptemp}
\end{subfigure}
\begin{minipage}[t]{.05\textwidth}
\hspace{3mm}
\end{minipage}
\begin{minipage}[t]{.4\textwidth}

\caption{Spatial correlations in the presence of a interface with $f=0.05$. DSMC results are indicated by the red diamonds. FHD results using a master equation for the interface are given by the black squares. In (\subref{fig005:velrho}), the correlation is shown between the velocity fluctuations at location $x^*$ and the density fluctuations at every other cell. The correlation between temperature fluctuations at $x^*$ and the density fluctuations in other cells is shown in (\subref{fig005:temprho}), and the temperature-temperature fluctuations are shown in (\subref{fig005:temptemp}). Each plot was generated using $S=1.6\times 10^{9}$ samples. Better agreement is observed than in Fig.~\ref{fig:comp01}, as the effusion condition is better met in this case. }\label{fig:comp005}
\end{minipage}

\end{figure}

In Figs.~\ref{fig:comp01} and \ref{fig:comp005}, the same correlations are shown in the presence of interfaces with effusion probability $f=0.1$ and $f=0.05$, respectively. In the FHD simulations, the master equation method of Section \ref{sec:masterimplement} has been used to simulate the interface. For the $f=0.05$ case, subject to the same caveats as above, good agreement is observed between the FHD and DSMC results in most of the domain, with some scatter in the FHD results in the vicinity of the interface. In the $f=0.1$ comparison, a small systematic difference is observed between the two sets of results, particularly visible in Figs.~\ref{fig01:velrho} and \ref{fig01:temptemp} -- this represents the breakdown of the effusion condition.

\begin{figure}
\centering

\begin{subfigure}[t]{.45\textwidth}
\centering
\includegraphics[width=\linewidth]{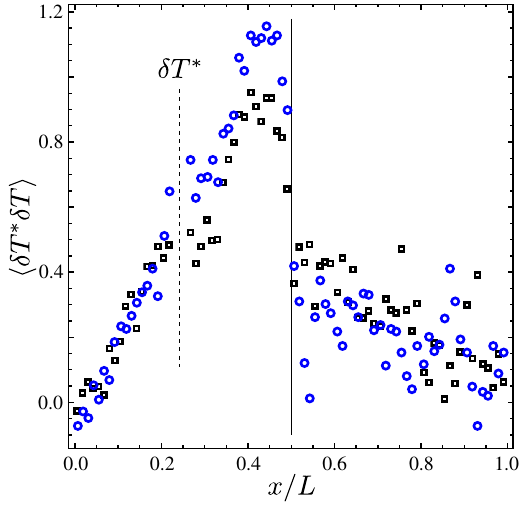}
        \caption{}\label{fig:compa}
\end{subfigure}
\begin{subfigure}[t]{.45\textwidth}
\centering
\includegraphics[width=\linewidth]{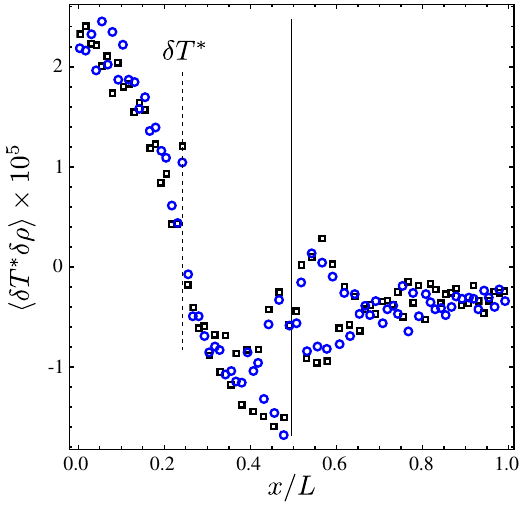}
\caption{}\label{fig:compb}
\end{subfigure}

\medskip

\begin{subfigure}[t]{.45\textwidth}
\centering
\includegraphics[width=\linewidth]{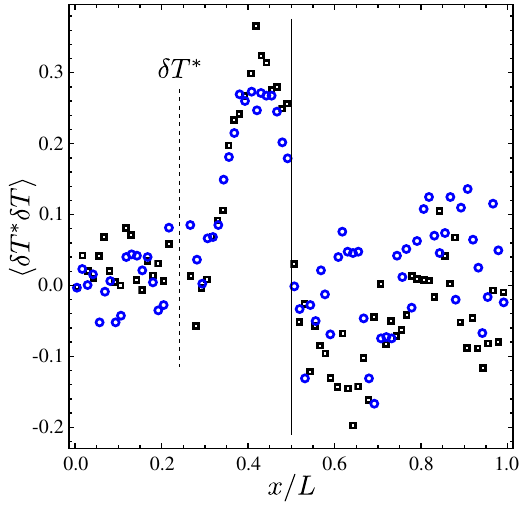}
        \caption{}\label{fig:compc}
\end{subfigure}
\begin{subfigure}[t]{.45\textwidth}
\centering
\includegraphics[width=\linewidth]{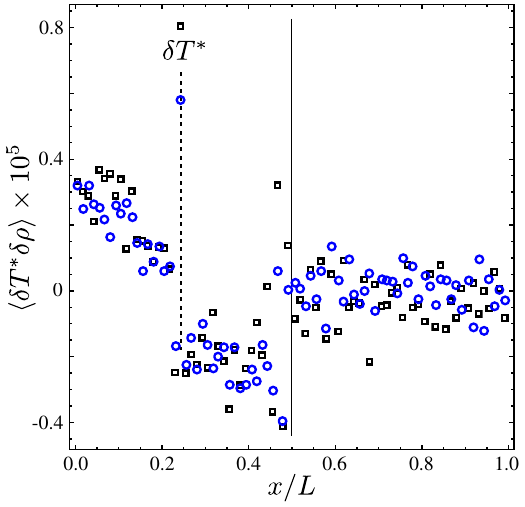}
\caption{}\label{fig:compd}
\end{subfigure}
\caption{Comparison of master equation (black squares) and Langevin (blue circles) methods for implementing interface. In the top row the case $f=0.1$ is shown, using the parameters given in Table~\ref{table:parameters}. In (\subref{fig:compa}), the correlation is shown between the temperature fluctuations at location $x^*$ and the temperature fluctuations at every other cell. The correlation between temperature fluctuations at $x^*$ and the density fluctuations in other cells is shown in (\subref{fig:compb}). Each plot was generated using $S=1.6\times 10^{9}$ samples. The bottom row, (\subref{fig:compc}) and (\subref{fig:compd}), displays the same results for the case $f=0.00625$. Each plot was generated using $S=3.2\times 10^{9}$ samples. While the $f=0.1$ results in the top row are separately compared to DSMC in Fig.~\ref{fig:comp01}, this has not been done for the $f=0.00625$ results in the bottom row due to the computational expense of the DSMC simulations.}\label{fig:complang}
\end{figure}

In Fig.~\ref{fig:complang} the master equation and Langevin approaches for modeling the interface, given in sections~\ref{sec:masterimplement} and \ref{sec:langevinimplement}, are compared. The correlations $\langle \delta T^* \delta \rho \rangle$ and $\langle \delta T^* \delta T \rangle$ are shown for the cases $f=0.1$ and $f=0.00625$. Excellent agreement is observed, demonstrating the equivalence of the two approaches.

As can be seen from Figs.~\ref{fig:compnomembrane}-\ref{fig:comp005}, the general impact of the interface is to reduce the magnitude of long range correlations in the gas. The value $\langle \delta u_x^* \delta \rho \rangle$ approaches zero for $x/L > 0.5$ because the interface does not conserve momentum. Interestingly, a new peak in $\langle \delta T^* \delta T \rangle$ develops immediately to the left of the membrane -- this behavior is borne out for lower values of $f$ not shown here. This distortion is also observed in a simple Fourier model, with the interface modeled as a region of low thermal conductivity.

\begin{figure}
\centering

\begin{subfigure}[t]{.45\textwidth}
\centering
\includegraphics[width=\linewidth]{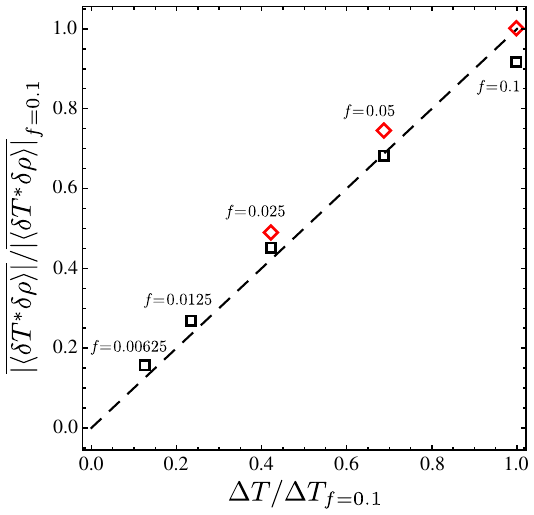}
        \caption{}\label{fig:lefttemprho}
\end{subfigure}
\begin{subfigure}[t]{.45\textwidth}
\centering
\includegraphics[width=\linewidth]{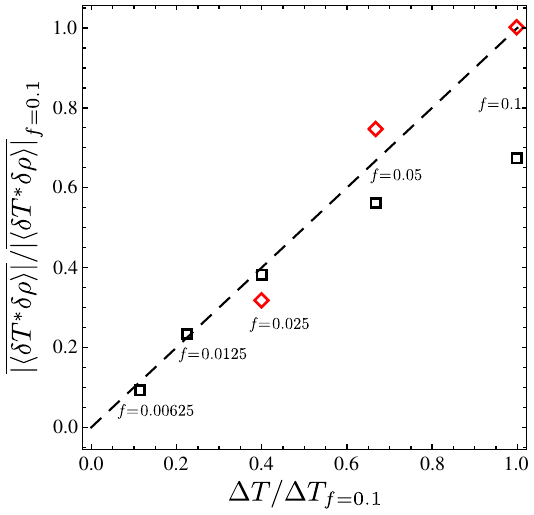}
        \caption{}\label{fig:righttemprho}
\end{subfigure}

\medskip

\begin{subfigure}[t]{.45\textwidth}
\centering
\vspace{0pt}
\includegraphics[width=\linewidth]{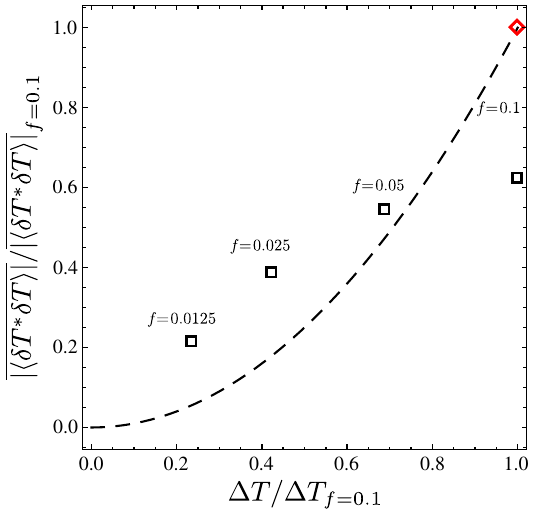}
        \caption{}\label{fig:lefttemptemp}
\end{subfigure}
\begin{subfigure}[t]{.45\textwidth}
\centering
\vspace{0pt}
\includegraphics[width=\linewidth]{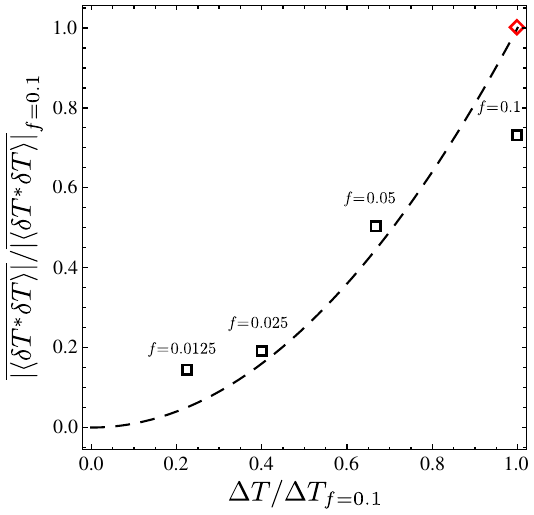}
        \caption{}\label{fig:righttemptemp}
\end{subfigure}

\caption{Response of spatial correlations (vertical axis) to changing temperature gradient (horizontal axis). Plots on the left and right show measurements from the left and right hand sides of the interface, respectively. In the top row, the value $\overline{|\langle \delta T^* \delta \rho \rangle|}$ indicates that the absolute value of the correlation has been averaged over the domain on that side of the interface, i.e., $0<x<L/2$ for (\subref{fig:lefttemprho}), and $L/2<x<L$ for (\subref{fig:righttemprho}). The red diamonds show DSMC results, and the black squares indicate FHD. Both the correlations and the gradients have been normalized by the values measured at $f=0.1$, with the DSMC solution being used for the correlations. The DSMC results were all obtained with a sample size $S=1.6\times 10^9$, while $S=8\times10^9$ for the FHD results.}\label{fig:gradscale}
\end{figure}

Aside from these features, the reduction in magnitude appears to arise principally from the change in temperature gradient. In Fig.~\ref{fig:fullgrad}, a reduction in the gradient is visible with decreasing $f$, where the interface behaves increasingly like an adiabatic surface. In Fig.~\ref{fig:gradscale}, the magnitude of correlations are shown as a function of the gradient. The top row shows measurements of the value $\overline{|\langle \delta T^* \delta \rho \rangle|}$, where the overline indicates the value has been smoothed by averaging over a spatial interval (see figure caption).
Fig.~\ref{fig:lefttemprho} shows measurements from the left hand side of the interface, while Fig.~\ref{fig:righttemprho} shows the values from the right. In the bottom row, Figs.~\ref{fig:lefttemptemp} and Fig.~\ref{fig:righttemptemp} show the left and right side measurements of $\overline{|\langle \delta T^* \delta T \rangle|}$. The results displayed are taken from DSMC and FHD simulations in the range $0.00625 \leq f \leq 0.1$, and the subsets shown in each plot are those where a statistically converged solution was obtained. Where both DSMC and FHD results are shown, the DSMC solution should be taken to be the more accurate; we have included the FHD results to illustrate the divergence of the solutions due to the violation of the effusion condition. For lower values of $f$, the magnitude of the correlations becomes smaller relative to equilibrium noise, requiring a larger sample size to obtain a result. FHD results were obtainable in this regime because this approach exhibits greater computational speed; this is discussed further in Section~\ref{sec:performance}.

In Figs.~\ref{fig:lefttemprho} and \ref{fig:righttemprho}, $\langle \delta T^* \delta \rho \rangle$ exhibits a linear response with respect to gradient; this agrees with the findings of Refs.~\citenum{Mans1} and \citenum{Nico1}, where a linear response to the temperature gradient is observed in the absence of a membrane. In the same references, a quadratic response is shown in $\langle \delta T^* \delta T \rangle$. This appears to be supported by Fig.~\ref{fig:righttemptemp}, in Fig.~\ref{fig:lefttemptemp} measurement of this effect is confounded by the appearance of the additional peak near the surface of the interface, discussed above.
Finally, it is clear from Figs.~\ref{fig:righttemprho} and \ref{fig:righttemptemp} that the fluctuations on opposite sides of the interface remain correlated even for small values of $f$.

%
%
%
%
%

\subsection{Comparative performance}\label{sec:performance}
The cell size used to make the comparisons shown in Figs.~\ref{fig:compnomembrane}-\ref{fig:comp005}, given in Table~\ref{table:parameters}, has been selected based on the kinetic length scale that constrains the DSMC simulations. The time step has been selected to minimize the chance of fluctuations causing negative density or energy values in the FHD simulations - these may also be avoided by increasing the cell size, which is not bound by the kinetic scale as in the DSMC simulations.

\begin{figure}
\centering

\begin{subfigure}[t]{.45\textwidth}
\centering
\includegraphics[width=\linewidth]{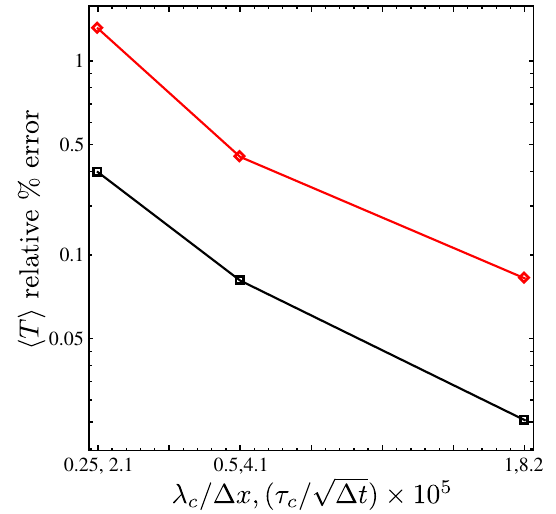}
        \caption{}\label{fig:conva}
\end{subfigure}
\begin{subfigure}[t]{.45\textwidth}
\centering
\includegraphics[width=0.965\linewidth]{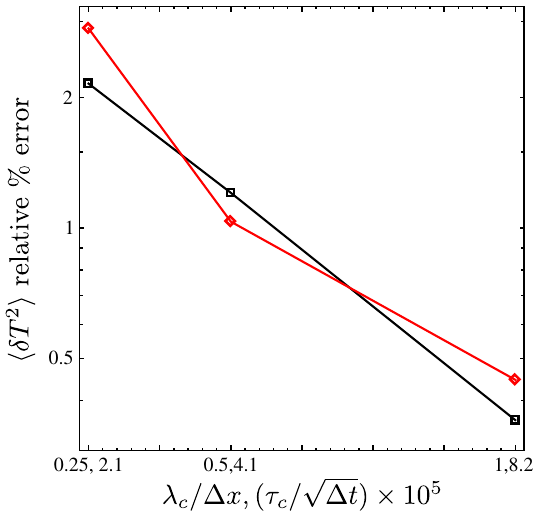}
\caption{}\label{fig:convb}
\end{subfigure}
\medskip

\begin{minipage}[t]{.95\textwidth}

\caption{Convergence of temperature mean and variance as a function of cell size an time step.  FHD results using the master equation approach for the interface are shown in black, the DSMC results are shown in red. Relative absolute error is calculated with respect to the cell size and time steps used in Figs.~\ref{fig:compnomembrane}-\ref{fig:comp005}, and shown in Table~\ref{table:parameters}. The errors shown are averaged over all cells in the domain. The changes in time step and cell size have been selected such that the diffusive stability value, $C\sim\Delta t / \Delta x^2$, remains constant. The time step is expressed relative to the hard sphere mean collision time, $\tau_\mathrm{c}$, and the cell size is given in terms of the hard sphere mean free path, $\lambda_\mathrm{c}$.}\label{fig:conv}
\end{minipage}

\end{figure}

In Fig.~\ref{fig:conv}, the effect of changing cell
size and time step is compared between the two methods, using the case $f=0.1$. To maintain a constant diffusive stability value, $C\sim \Delta t / \Delta x^2$, for every doubling of the cell size the time step is increased by a factor of four. Due to the very small time steps used, the error appears to be dominated by the cell size, with both methods showing the expected quadratic convergence. 
The FHD algorithm is deterministically third order in time and second order in space and weakly second order for stochastic systems; DSMC is at best second order. We note also that, depending on the simulation parameters, one FHD time step executes approximately ten to thirty times faster than the equivalent DSMC time step.

\section{Concluding Remarks}\label{sec:Conclusion}
In this article, we have investigated the effect of a simple transpiration interface on long range hydrodynamic correlations in a dilute gas, where a non-equilibrium steady state has been induced by a temperature gradient. Several qualitative differences in the correlations were observed, however the principal effect of the interface was to reduce their magnitude proportionally to the effusion probability, $f$. This outcome can largely be explained by changes in the temperature gradient induced by the interface. We also observe a distortion of the $\langle \delta T^* \delta T \rangle$ correlations, specifically the appearance of a new peak located near the interface.

The investigation used two distinct numerical techniques, DSMC and FHD, which were found to be in agreement for appropriate values of $f$; for $f \gtrsim 0.1$, non-continuum effects reduce the accuracy of the FHD method. The FHD simulations used both master equation and Langevin approaches for modeling the interface, with the two methods producing equivalent results. This validates the use of FHD for interfaces with low effusion probabilities, giving a computationally faster alternative to DSMC.

There are a number of potential extensions to the work presented here.  Currently, both methods used to model the interface in the FHD simulations assume $J_{\mathbfcal{P}}=0$, i.e., there is zero momentum flux. Future extensions to the interface model could include
momentum fluctuations around a zero mean, 
a more realistic model of thermal processes
and incorporation of multicomponent effects such as species dependent permeability.

These developments would provide the starting point for numerical modeling of a range of membrane behaviors. More complex models could include models for transport of ions through membranes and the incorporation of chemical reactions that enable active transport. These capabilities would enable the methodology be used to model transport in biological membranes and transport process important to artificial photosynthesis.


%
%

%

\begin{acknowledgments}
The authors wish to acknowledge many years of fruitful discussions with Graeme Bird.
This work was supported by the
U.S.~Department of Energy, Office of Science,
Office of Advanced Scientific Computing Research,
Applied Mathematics Program under contract DE-AC02-05CH11231.
This research used resources of the National Energy Research Scientific Computing Center, a DOE Office of Science User Facility supported by the Office of Science of the U.S. Department of Energy under Contract No. DE-AC02-05CH11231.
\end{acknowledgments}

\clearpage
\bibliography{references.bib}

\providecommand{\noopsort}[1]{}\providecommand{\singleletter}[1]{#1}%
\begin{thebibliography}{51}%
\makeatletter
\providecommand \@ifxundefined [1]{%
 \@ifx{#1\undefined}
}%
\providecommand \@ifnum [1]{%
 \ifnum #1\expandafter \@firstoftwo
 \else \expandafter \@secondoftwo
 \fi
}%
\providecommand \@ifx [1]{%
 \ifx #1\expandafter \@firstoftwo
 \else \expandafter \@secondoftwo
 \fi
}%
\providecommand \natexlab [1]{#1}%
\providecommand \enquote  [1]{``#1''}%
\providecommand \bibnamefont  [1]{#1}%
\providecommand \bibfnamefont [1]{#1}%
\providecommand \citenamefont [1]{#1}%
\providecommand \href@noop [0]{\@secondoftwo}%
\providecommand \href [0]{\begingroup \@sanitize@url \@href}%
\providecommand \@href[1]{\@@startlink{#1}\@@href}%
\providecommand \@@href[1]{\endgroup#1\@@endlink}%
\providecommand \@sanitize@url [0]{\catcode `\\12\catcode `\$12\catcode
  `\&12\catcode `\#12\catcode `\^12\catcode `\_12\catcode `\%12\relax}%
\providecommand \@@startlink[1]{}%
\providecommand \@@endlink[0]{}%
\providecommand \url  [0]{\begingroup\@sanitize@url \@url }%
\providecommand \@url [1]{\endgroup\@href {#1}{\urlprefix }}%
\providecommand \urlprefix  [0]{URL }%
\providecommand \Eprint [0]{\href }%
\providecommand \doibase [0]{http://dx.doi.org/}%
\providecommand \selectlanguage [0]{\@gobble}%
\providecommand \bibinfo  [0]{\@secondoftwo}%
\providecommand \bibfield  [0]{\@secondoftwo}%
\providecommand \translation [1]{[#1]}%
\providecommand \BibitemOpen [0]{}%
\providecommand \bibitemStop [0]{}%
\providecommand \bibitemNoStop [0]{.\EOS\space}%
\providecommand \EOS [0]{\spacefactor3000\relax}%
\providecommand \BibitemShut  [1]{\csname bibitem#1\endcsname}%
\let\auto@bib@innerbib\@empty
\bibitem [{\citenamefont {H\"anggi}\ and\ \citenamefont
  {Marchesoni}(2009)}]{HanggiBrownianMotors}%
  \BibitemOpen
  \bibfield  {author} {\bibinfo {author} {\bibfnamefont {P.}~\bibnamefont
  {H\"anggi}}\ and\ \bibinfo {author} {\bibfnamefont {F.}~\bibnamefont
  {Marchesoni}},\ }\bibfield  {title} {\enquote {\bibinfo {title} {Artificial
  brownian motors: Controlling transport on the nanoscale},}\ }\href {\doibase
  10.1103/RevModPhys.81.387} {\bibfield  {journal} {\bibinfo  {journal} {Rev.
  Mod. Phys.}\ }\textbf {\bibinfo {volume} {81}},\ \bibinfo {pages} {387--442}
  (\bibinfo {year} {2009})}\BibitemShut {NoStop}%
\bibitem [{\citenamefont {Mikhailov}\ and\ \citenamefont
  {Kapral}(2015)}]{Kapral2015}%
  \BibitemOpen
  \bibfield  {author} {\bibinfo {author} {\bibfnamefont {A.~S.}\ \bibnamefont
  {Mikhailov}}\ and\ \bibinfo {author} {\bibfnamefont {R.}~\bibnamefont
  {Kapral}},\ }\bibfield  {title} {\enquote {\bibinfo {title} {Hydrodynamic
  collective effects of active protein machines in solution and lipid
  bilayers},}\ }\href@noop {} {\bibfield  {journal} {\bibinfo  {journal}
  {Proceedings of the National Academy of Sciences}\ }\textbf {\bibinfo
  {volume} {112}},\ \bibinfo {pages} {E3639--E3644} (\bibinfo {year}
  {2015})}\BibitemShut {NoStop}%
\bibitem [{\citenamefont {Berne}\ and\ \citenamefont
  {Pecora}(2013)}]{berne2013dynamic}%
  \BibitemOpen
  \bibfield  {author} {\bibinfo {author} {\bibfnamefont {B.}~\bibnamefont
  {Berne}}\ and\ \bibinfo {author} {\bibfnamefont {R.}~\bibnamefont {Pecora}},\
  }\href {https://books.google.com/books?id=xg3CAgAAQBAJ} {\emph {\bibinfo
  {title} {Dynamic Light Scattering: With Applications to Chemistry, Biology,
  and Physics}}},\ Dover Books on Physics\ (\bibinfo  {publisher} {Dover
  Publications},\ \bibinfo {year} {2013})\BibitemShut {NoStop}%
\bibitem [{\citenamefont {Lemarchand}\ and\ \citenamefont
  {Nowakowski}(2004)}]{Lemarchand2}%
  \BibitemOpen
  \bibfield  {author} {\bibinfo {author} {\bibfnamefont {A.}~\bibnamefont
  {Lemarchand}}\ and\ \bibinfo {author} {\bibfnamefont {B.}~\bibnamefont
  {Nowakowski}},\ }\bibfield  {title} {\enquote {\bibinfo {title}
  {Fluctuation-induced and nonequilibrium-induced bifurcations in a
  thermochemical system},}\ }\href@noop {} {\bibfield  {journal} {\bibinfo
  {journal} {Molecular Simulation}\ }\textbf {\bibinfo {volume} {30}},\
  \bibinfo {pages} {773--780} (\bibinfo {year} {2004})}\BibitemShut {NoStop}%
\bibitem [{\citenamefont {Donev}\ \emph
  {et~al.}(2011{\natexlab{a}})\citenamefont {Donev}, \citenamefont {Bell},
  \citenamefont {de~La~Fuente},\ and\ \citenamefont {Garcia}}]{Done3}%
  \BibitemOpen
  \bibfield  {author} {\bibinfo {author} {\bibfnamefont {A.}~\bibnamefont
  {Donev}}, \bibinfo {author} {\bibfnamefont {J.~B.}\ \bibnamefont {Bell}},
  \bibinfo {author} {\bibfnamefont {A.}~\bibnamefont {de~La~Fuente}}, \ and\
  \bibinfo {author} {\bibfnamefont {A.~L.}\ \bibnamefont {Garcia}},\ }\bibfield
   {title} {\enquote {\bibinfo {title} {Diffusive transport by thermal velocity
  fluctuations},}\ }\href@noop {} {\bibfield  {journal} {\bibinfo  {journal}
  {Physical review letters}\ }\textbf {\bibinfo {volume} {106}},\ \bibinfo
  {pages} {204501} (\bibinfo {year} {2011}{\natexlab{a}})}\BibitemShut
  {NoStop}%
\bibitem [{\citenamefont {{Vailati}}\ \emph {et~al.}(2011)\citenamefont
  {{Vailati}}, \citenamefont {{Cerbino}}, \citenamefont {{Mazzoni}},
  \citenamefont {{Takacs}}, \citenamefont {{Cannell}},\ and\ \citenamefont
  {{Giglio}}}]{Vailati1}%
  \BibitemOpen
  \bibfield  {author} {\bibinfo {author} {\bibfnamefont {A.}~\bibnamefont
  {{Vailati}}}, \bibinfo {author} {\bibfnamefont {R.}~\bibnamefont
  {{Cerbino}}}, \bibinfo {author} {\bibfnamefont {S.}~\bibnamefont
  {{Mazzoni}}}, \bibinfo {author} {\bibfnamefont {C.~J.}\ \bibnamefont
  {{Takacs}}}, \bibinfo {author} {\bibfnamefont {D.~S.}\ \bibnamefont
  {{Cannell}}}, \ and\ \bibinfo {author} {\bibfnamefont {M.}~\bibnamefont
  {{Giglio}}},\ }\bibfield  {title} {\enquote {\bibinfo {title} {{Fractal
  fronts of diffusion in microgravity}},}\ }\href@noop {} {\bibfield  {journal}
  {\bibinfo  {journal} {Nature Communications}\ }\textbf {\bibinfo {volume}
  {2}},\ \bibinfo {eid} {290} (\bibinfo {year} {2011})}\BibitemShut {NoStop}%
\bibitem [{\citenamefont {Morgado}, \citenamefont {Nowakowski},\ and\
  \citenamefont {Lemarchand}(2018)}]{Lemarchand1}%
  \BibitemOpen
  \bibfield  {author} {\bibinfo {author} {\bibfnamefont {G.}~\bibnamefont
  {Morgado}}, \bibinfo {author} {\bibfnamefont {B.}~\bibnamefont {Nowakowski}},
  \ and\ \bibinfo {author} {\bibfnamefont {A.}~\bibnamefont {Lemarchand}},\
  }\bibfield  {title} {\enquote {\bibinfo {title} {Scaling of submicrometric
  turing patterns in concentrated growing systems},}\ }\href@noop {} {\bibfield
   {journal} {\bibinfo  {journal} {Phys. Rev. E}\ }\textbf {\bibinfo {volume}
  {98}},\ \bibinfo {pages} {032213} (\bibinfo {year} {2018})}\BibitemShut
  {NoStop}%
\bibitem [{\citenamefont {Gallis}\ \emph {et~al.}(2015)\citenamefont {Gallis},
  \citenamefont {Koehler}, \citenamefont {Torczynski},\ and\ \citenamefont
  {Plimpton}}]{Gallis2015}%
  \BibitemOpen
  \bibfield  {author} {\bibinfo {author} {\bibfnamefont {M.~A.}\ \bibnamefont
  {Gallis}}, \bibinfo {author} {\bibfnamefont {T.~P.}\ \bibnamefont {Koehler}},
  \bibinfo {author} {\bibfnamefont {J.~R.}\ \bibnamefont {Torczynski}}, \ and\
  \bibinfo {author} {\bibfnamefont {S.~J.}\ \bibnamefont {Plimpton}},\
  }\bibfield  {title} {\enquote {\bibinfo {title} {Direct simulation monte
  carlo investigation of the richtmyer-meshkov instability},}\ }\href@noop {}
  {\bibfield  {journal} {\bibinfo  {journal} {Physics of Fluids}\ }\textbf
  {\bibinfo {volume} {27}},\ \bibinfo {pages} {084105} (\bibinfo {year}
  {2015})}\BibitemShut {NoStop}%
\bibitem [{\citenamefont {Kirkpatrick}, \citenamefont {Ortiz~de Z\'arate},\
  and\ \citenamefont {Sengers}(2015)}]{CasimirPrl2015}%
  \BibitemOpen
  \bibfield  {author} {\bibinfo {author} {\bibfnamefont {T.~R.}\ \bibnamefont
  {Kirkpatrick}}, \bibinfo {author} {\bibfnamefont {J.~M.}\ \bibnamefont
  {Ortiz~de Z\'arate}}, \ and\ \bibinfo {author} {\bibfnamefont {J.~V.}\
  \bibnamefont {Sengers}},\ }\bibfield  {title} {\enquote {\bibinfo {title}
  {Nonequilibrium casimir-like forces in liquid mixtures},}\ }\href@noop {}
  {\bibfield  {journal} {\bibinfo  {journal} {Phys. Rev. Lett.}\ }\textbf
  {\bibinfo {volume} {115}},\ \bibinfo {pages} {035901} (\bibinfo {year}
  {2015})}\BibitemShut {NoStop}%
\bibitem [{\citenamefont {Frenkel}\ and\ \citenamefont {Smit}(2001)}]{Fren1}%
  \BibitemOpen
  \bibfield  {author} {\bibinfo {author} {\bibfnamefont {D.}~\bibnamefont
  {Frenkel}}\ and\ \bibinfo {author} {\bibfnamefont {B.}~\bibnamefont {Smit}},\
  }\href {https://books.google.com/books?id=5qTzldS9ROIC} {\emph {\bibinfo
  {title} {Understanding Molecular Simulation: From Algorithms to
  Applications}}},\ Computational science series\ (\bibinfo  {publisher}
  {Elsevier Science},\ \bibinfo {year} {2001})\BibitemShut {NoStop}%
\bibitem [{\citenamefont {Bird}(1994)}]{Bird1}%
  \BibitemOpen
  \bibfield  {author} {\bibinfo {author} {\bibfnamefont {G.~A.}\ \bibnamefont
  {Bird}},\ }\href@noop {} {\emph {\bibinfo {title} {Molecular Gas Dynamics and
  the Direct Simulation of Gas Flows}}}\ (\bibinfo  {publisher} {Oxford
  Engineering Science Series},\ \bibinfo {year} {1994})\BibitemShut {NoStop}%
\bibitem [{\citenamefont {Bird}(2013)}]{Bird8}%
  \BibitemOpen
  \bibfield  {author} {\bibinfo {author} {\bibfnamefont {G.}~\bibnamefont
  {Bird}},\ }\href {https://books.google.com/books?id=0YdnngEACAAJ} {\emph
  {\bibinfo {title} {The DSMC Method}}}\ (\bibinfo  {publisher} {CreateSpace
  Independent Publishing Platform},\ \bibinfo {year} {2013})\BibitemShut
  {NoStop}%
\bibitem [{\citenamefont {Wagner}(1992)}]{Wagn1}%
  \BibitemOpen
  \bibfield  {author} {\bibinfo {author} {\bibfnamefont {W.}~\bibnamefont
  {Wagner}},\ }\bibfield  {title} {\enquote {\bibinfo {title} {A convergence
  proof for {B}ird's direct simulation {M}onte {C}arlo method for the
  {B}oltzmann equation},}\ }\href@noop {} {\bibfield  {journal} {\bibinfo
  {journal} {Journal of Statistical Physics}\ }\textbf {\bibinfo {volume}
  {66}},\ \bibinfo {pages} {1011--1044} (\bibinfo {year} {1992})}\BibitemShut
  {NoStop}%
\bibitem [{\citenamefont {Sone}(2000)}]{Sone1}%
  \BibitemOpen
  \bibfield  {author} {\bibinfo {author} {\bibfnamefont {Y.}~\bibnamefont
  {Sone}},\ }\href@noop {} {\emph {\bibinfo {title} {Kinetic Theory and Fluid
  Dynamics}}}\ (\bibinfo  {publisher} {Birkh{\"a}user},\ \bibinfo {year}
  {2000})\BibitemShut {NoStop}%
\bibitem [{\citenamefont {Landau}\ and\ \citenamefont
  {Lifshitz}(1959)}]{Land2}%
  \BibitemOpen
  \bibfield  {author} {\bibinfo {author} {\bibfnamefont {L.}~\bibnamefont
  {Landau}}\ and\ \bibinfo {author} {\bibfnamefont {E.}~\bibnamefont
  {Lifshitz}},\ }\href@noop {} {\emph {\bibinfo {title} {Course of theoretical
  physics. Vol. 6: {F}luid mechanics}}}\ (\bibinfo  {publisher} {London},\
  \bibinfo {year} {1959})\BibitemShut {NoStop}%
\bibitem [{\citenamefont {de~Zarate}\ and\ \citenamefont
  {Sengers}(2006)}]{Zara1}%
  \BibitemOpen
  \bibfield  {author} {\bibinfo {author} {\bibfnamefont {J.}~\bibnamefont
  {de~Zarate}}\ and\ \bibinfo {author} {\bibfnamefont {J.}~\bibnamefont
  {Sengers}},\ }\href {https://books.google.com/books?id=4sx3CVMreJMC} {\emph
  {\bibinfo {title} {Hydrodynamic Fluctuations in Fluids and Fluid Mixtures}}}\
  (\bibinfo  {publisher} {Elsevier Science},\ \bibinfo {year}
  {2006})\BibitemShut {NoStop}%
\bibitem [{\citenamefont {Segr\`e}\ \emph {et~al.}(1992)\citenamefont
  {Segr\`e}, \citenamefont {Gammon}, \citenamefont {Sengers},\ and\
  \citenamefont {Law}}]{LightFhd1994}%
  \BibitemOpen
  \bibfield  {author} {\bibinfo {author} {\bibfnamefont {P.~N.}\ \bibnamefont
  {Segr\`e}}, \bibinfo {author} {\bibfnamefont {R.~W.}\ \bibnamefont {Gammon}},
  \bibinfo {author} {\bibfnamefont {J.~V.}\ \bibnamefont {Sengers}}, \ and\
  \bibinfo {author} {\bibfnamefont {B.~M.}\ \bibnamefont {Law}},\ }\bibfield
  {title} {\enquote {\bibinfo {title} {Rayleigh scattering in a liquid far from
  thermal equilibrium},}\ }\href@noop {} {\bibfield  {journal} {\bibinfo
  {journal} {Phys. Rev. A}\ }\textbf {\bibinfo {volume} {45}},\ \bibinfo
  {pages} {714--724} (\bibinfo {year} {1992})}\BibitemShut {NoStop}%
\bibitem [{\citenamefont {Takacs}\ \emph {et~al.}(2011)\citenamefont {Takacs},
  \citenamefont {Vailati}, \citenamefont {Cerbino}, \citenamefont {Mazzoni},
  \citenamefont {Giglio},\ and\ \citenamefont {Cannell}}]{ShadowFhd2011}%
  \BibitemOpen
  \bibfield  {author} {\bibinfo {author} {\bibfnamefont {C.~J.}\ \bibnamefont
  {Takacs}}, \bibinfo {author} {\bibfnamefont {A.}~\bibnamefont {Vailati}},
  \bibinfo {author} {\bibfnamefont {R.}~\bibnamefont {Cerbino}}, \bibinfo
  {author} {\bibfnamefont {S.}~\bibnamefont {Mazzoni}}, \bibinfo {author}
  {\bibfnamefont {M.}~\bibnamefont {Giglio}}, \ and\ \bibinfo {author}
  {\bibfnamefont {D.~S.}\ \bibnamefont {Cannell}},\ }\bibfield  {title}
  {\enquote {\bibinfo {title} {Thermal fluctuations in a layer of liquid
  ${\mathrm{cs}}_{2}$ subjected to temperature gradients with and without the
  influence of gravity},}\ }\href@noop {} {\bibfield  {journal} {\bibinfo
  {journal} {Phys. Rev. Lett.}\ }\textbf {\bibinfo {volume} {106}},\ \bibinfo
  {pages} {244502} (\bibinfo {year} {2011})}\BibitemShut {NoStop}%
\bibitem [{\citenamefont {Bell}, \citenamefont {Garcia},\ and\ \citenamefont
  {Williams}(2007)}]{Bell1}%
  \BibitemOpen
  \bibfield  {author} {\bibinfo {author} {\bibfnamefont {J.~B.}\ \bibnamefont
  {Bell}}, \bibinfo {author} {\bibfnamefont {A.~L.}\ \bibnamefont {Garcia}}, \
  and\ \bibinfo {author} {\bibfnamefont {S.~A.}\ \bibnamefont {Williams}},\
  }\bibfield  {title} {\enquote {\bibinfo {title} {Numerical methods for the
  stochastic {L}andau-{L}ifshitz {N}avier-{S}tokes equations},}\ }\href@noop {}
  {\bibfield  {journal} {\bibinfo  {journal} {Physical Review E}\ }\textbf
  {\bibinfo {volume} {76}},\ \bibinfo {pages} {016708} (\bibinfo {year}
  {2007})}\BibitemShut {NoStop}%
\bibitem [{\citenamefont {Donev}\ \emph
  {et~al.}(2010{\natexlab{a}})\citenamefont {Donev}, \citenamefont
  {Vanden-Eijnden}, \citenamefont {Garcia},\ and\ \citenamefont
  {Bell}}]{Done1}%
  \BibitemOpen
  \bibfield  {author} {\bibinfo {author} {\bibfnamefont {A.}~\bibnamefont
  {Donev}}, \bibinfo {author} {\bibfnamefont {E.}~\bibnamefont
  {Vanden-Eijnden}}, \bibinfo {author} {\bibfnamefont {A.}~\bibnamefont
  {Garcia}}, \ and\ \bibinfo {author} {\bibfnamefont {J.}~\bibnamefont
  {Bell}},\ }\bibfield  {title} {\enquote {\bibinfo {title} {On the accuracy of
  finite-volume schemes for fluctuating hydrodynamics},}\ }\href@noop {}
  {\bibfield  {journal} {\bibinfo  {journal} {Communications in Applied
  Mathematics and Computational Science}\ }\textbf {\bibinfo {volume} {5}},\
  \bibinfo {pages} {149--197} (\bibinfo {year}
  {2010}{\natexlab{a}})}\BibitemShut {NoStop}%
\bibitem [{\citenamefont {Balboa}\ \emph {et~al.}(2012)\citenamefont {Balboa},
  \citenamefont {Bell}, \citenamefont {Delgado-Buscalioni}, \citenamefont
  {Donev}, \citenamefont {Fai}, \citenamefont {Griffith},\ and\ \citenamefont
  {Peskin}}]{Balb1}%
  \BibitemOpen
  \bibfield  {author} {\bibinfo {author} {\bibfnamefont {F.}~\bibnamefont
  {Balboa}}, \bibinfo {author} {\bibfnamefont {J.~B.}\ \bibnamefont {Bell}},
  \bibinfo {author} {\bibfnamefont {R.}~\bibnamefont {Delgado-Buscalioni}},
  \bibinfo {author} {\bibfnamefont {A.}~\bibnamefont {Donev}}, \bibinfo
  {author} {\bibfnamefont {T.~G.}\ \bibnamefont {Fai}}, \bibinfo {author}
  {\bibfnamefont {B.~E.}\ \bibnamefont {Griffith}}, \ and\ \bibinfo {author}
  {\bibfnamefont {C.~S.}\ \bibnamefont {Peskin}},\ }\bibfield  {title}
  {\enquote {\bibinfo {title} {Staggered schemes for fluctuating
  hydrodynamics},}\ }\href@noop {} {\bibfield  {journal} {\bibinfo  {journal}
  {Multiscale Modeling \& Simulation}\ }\textbf {\bibinfo {volume} {10}},\
  \bibinfo {pages} {1369--1408} (\bibinfo {year} {2012})}\BibitemShut {NoStop}%
\bibitem [{\citenamefont {Bell}, \citenamefont {Garcia},\ and\ \citenamefont
  {Williams}(2010)}]{Bell2}%
  \BibitemOpen
  \bibfield  {author} {\bibinfo {author} {\bibfnamefont {J.~B.}\ \bibnamefont
  {Bell}}, \bibinfo {author} {\bibfnamefont {A.~L.}\ \bibnamefont {Garcia}}, \
  and\ \bibinfo {author} {\bibfnamefont {S.~A.}\ \bibnamefont {Williams}},\
  }\bibfield  {title} {\enquote {\bibinfo {title} {Computational fluctuating
  fluid dynamics},}\ }\href@noop {} {\bibfield  {journal} {\bibinfo  {journal}
  {ESAIM: Mathematical Modelling and Numerical Analysis}\ }\textbf {\bibinfo
  {volume} {44}},\ \bibinfo {pages} {1085--1105} (\bibinfo {year}
  {2010})}\BibitemShut {NoStop}%
\bibitem [{\citenamefont {Donev}\ \emph
  {et~al.}(2011{\natexlab{b}})\citenamefont {Donev}, \citenamefont {Bell},
  \citenamefont {De~la Fuente},\ and\ \citenamefont {Garcia}}]{Done4}%
  \BibitemOpen
  \bibfield  {author} {\bibinfo {author} {\bibfnamefont {A.}~\bibnamefont
  {Donev}}, \bibinfo {author} {\bibfnamefont {J.~B.}\ \bibnamefont {Bell}},
  \bibinfo {author} {\bibfnamefont {A.}~\bibnamefont {De~la Fuente}}, \ and\
  \bibinfo {author} {\bibfnamefont {A.~L.}\ \bibnamefont {Garcia}},\ }\bibfield
   {title} {\enquote {\bibinfo {title} {Enhancement of diffusive transport by
  non-equilibrium thermal fluctuations},}\ }\href@noop {} {\bibfield  {journal}
  {\bibinfo  {journal} {Journal of Statistical Mechanics: Theory and
  Experiment}\ }\textbf {\bibinfo {volume} {2011}},\ \bibinfo {pages} {P06014}
  (\bibinfo {year} {2011}{\natexlab{b}})}\BibitemShut {NoStop}%
\bibitem [{\citenamefont {Balakrishnan}\ \emph {et~al.}(2014)\citenamefont
  {Balakrishnan}, \citenamefont {Garcia}, \citenamefont {Donev},\ and\
  \citenamefont {Bell}}]{Bala1}%
  \BibitemOpen
  \bibfield  {author} {\bibinfo {author} {\bibfnamefont {K.}~\bibnamefont
  {Balakrishnan}}, \bibinfo {author} {\bibfnamefont {A.~L.}\ \bibnamefont
  {Garcia}}, \bibinfo {author} {\bibfnamefont {A.}~\bibnamefont {Donev}}, \
  and\ \bibinfo {author} {\bibfnamefont {J.~B.}\ \bibnamefont {Bell}},\
  }\bibfield  {title} {\enquote {\bibinfo {title} {Fluctuating hydrodynamics of
  multispecies nonreactive mixtures},}\ }\href@noop {} {\bibfield  {journal}
  {\bibinfo  {journal} {Physical Review E}\ }\textbf {\bibinfo {volume} {89}},\
  \bibinfo {pages} {013017} (\bibinfo {year} {2014})}\BibitemShut {NoStop}%
\bibitem [{\citenamefont {Donev}\ \emph {et~al.}(2014)\citenamefont {Donev},
  \citenamefont {Nonaka}, \citenamefont {Sun}, \citenamefont {Fai},
  \citenamefont {Garcia},\ and\ \citenamefont {Bell}}]{Done5}%
  \BibitemOpen
  \bibfield  {author} {\bibinfo {author} {\bibfnamefont {A.}~\bibnamefont
  {Donev}}, \bibinfo {author} {\bibfnamefont {A.}~\bibnamefont {Nonaka}},
  \bibinfo {author} {\bibfnamefont {Y.}~\bibnamefont {Sun}}, \bibinfo {author}
  {\bibfnamefont {T.}~\bibnamefont {Fai}}, \bibinfo {author} {\bibfnamefont
  {A.}~\bibnamefont {Garcia}}, \ and\ \bibinfo {author} {\bibfnamefont
  {J.}~\bibnamefont {Bell}},\ }\bibfield  {title} {\enquote {\bibinfo {title}
  {Low {M}ach number fluctuating hydrodynamics of diffusively mixing fluids},}\
  }\href@noop {} {\bibfield  {journal} {\bibinfo  {journal} {Communications in
  Applied Mathematics and Computational Science}\ }\textbf {\bibinfo {volume}
  {9}},\ \bibinfo {pages} {47--105} (\bibinfo {year} {2014})}\BibitemShut
  {NoStop}%
\bibitem [{\citenamefont {Nonaka}\ \emph {et~al.}(2015)\citenamefont {Nonaka},
  \citenamefont {Sun}, \citenamefont {Bell},\ and\ \citenamefont
  {Donev}}]{Nona1}%
  \BibitemOpen
  \bibfield  {author} {\bibinfo {author} {\bibfnamefont {A.}~\bibnamefont
  {Nonaka}}, \bibinfo {author} {\bibfnamefont {Y.}~\bibnamefont {Sun}},
  \bibinfo {author} {\bibfnamefont {J.}~\bibnamefont {Bell}}, \ and\ \bibinfo
  {author} {\bibfnamefont {A.}~\bibnamefont {Donev}},\ }\bibfield  {title}
  {\enquote {\bibinfo {title} {Low {M}ach number fluctuating hydrodynamics of
  binary liquid mixtures},}\ }\href@noop {} {\bibfield  {journal} {\bibinfo
  {journal} {Communications in Applied Mathematics and Computational Science}\
  }\textbf {\bibinfo {volume} {10}},\ \bibinfo {pages} {163--204} (\bibinfo
  {year} {2015})}\BibitemShut {NoStop}%
\bibitem [{\citenamefont {Donev}\ \emph {et~al.}(2015)\citenamefont {Donev},
  \citenamefont {Nonaka}, \citenamefont {Bhattacharjee}, \citenamefont
  {Garcia},\ and\ \citenamefont {Bell}}]{Done6}%
  \BibitemOpen
  \bibfield  {author} {\bibinfo {author} {\bibfnamefont {A.}~\bibnamefont
  {Donev}}, \bibinfo {author} {\bibfnamefont {A.}~\bibnamefont {Nonaka}},
  \bibinfo {author} {\bibfnamefont {A.~K.}\ \bibnamefont {Bhattacharjee}},
  \bibinfo {author} {\bibfnamefont {A.~L.}\ \bibnamefont {Garcia}}, \ and\
  \bibinfo {author} {\bibfnamefont {J.~B.}\ \bibnamefont {Bell}},\ }\bibfield
  {title} {\enquote {\bibinfo {title} {Low {M}ach number fluctuating
  hydrodynamics of multispecies liquid mixtures},}\ }\href@noop {} {\bibfield
  {journal} {\bibinfo  {journal} {Physics of Fluids}\ }\textbf {\bibinfo
  {volume} {27}},\ \bibinfo {pages} {037103} (\bibinfo {year}
  {2015})}\BibitemShut {NoStop}%
\bibitem [{\citenamefont {Chaudhri}\ \emph {et~al.}(2014)\citenamefont
  {Chaudhri}, \citenamefont {Bell}, \citenamefont {Garcia},\ and\ \citenamefont
  {Donev}}]{Chau1}%
  \BibitemOpen
  \bibfield  {author} {\bibinfo {author} {\bibfnamefont {A.}~\bibnamefont
  {Chaudhri}}, \bibinfo {author} {\bibfnamefont {J.~B.}\ \bibnamefont {Bell}},
  \bibinfo {author} {\bibfnamefont {A.~L.}\ \bibnamefont {Garcia}}, \ and\
  \bibinfo {author} {\bibfnamefont {A.}~\bibnamefont {Donev}},\ }\bibfield
  {title} {\enquote {\bibinfo {title} {Modeling multiphase flow using
  fluctuating hydrodynamics},}\ }\href@noop {} {\bibfield  {journal} {\bibinfo
  {journal} {Physical Review E}\ }\textbf {\bibinfo {volume} {90}},\ \bibinfo
  {pages} {033014} (\bibinfo {year} {2014})}\BibitemShut {NoStop}%
\bibitem [{\citenamefont {Bhattacharjee}\ \emph {et~al.}(2015)\citenamefont
  {Bhattacharjee}, \citenamefont {Balakrishnan}, \citenamefont {Garcia},
  \citenamefont {Bell},\ and\ \citenamefont {Donev}}]{Bhat2}%
  \BibitemOpen
  \bibfield  {author} {\bibinfo {author} {\bibfnamefont {A.~K.}\ \bibnamefont
  {Bhattacharjee}}, \bibinfo {author} {\bibfnamefont {K.}~\bibnamefont
  {Balakrishnan}}, \bibinfo {author} {\bibfnamefont {A.~L.}\ \bibnamefont
  {Garcia}}, \bibinfo {author} {\bibfnamefont {J.~B.}\ \bibnamefont {Bell}}, \
  and\ \bibinfo {author} {\bibfnamefont {A.}~\bibnamefont {Donev}},\ }\bibfield
   {title} {\enquote {\bibinfo {title} {Fluctuating hydrodynamics of
  multi-species reactive mixtures},}\ }\href@noop {} {\bibfield  {journal}
  {\bibinfo  {journal} {The Journal of chemical physics}\ }\textbf {\bibinfo
  {volume} {142}},\ \bibinfo {pages} {224107} (\bibinfo {year}
  {2015})}\BibitemShut {NoStop}%
\bibitem [{\citenamefont {P{\'e}raud}\ \emph {et~al.}(2016)\citenamefont
  {P{\'e}raud}, \citenamefont {Nonaka}, \citenamefont {Chaudhri}, \citenamefont
  {Bell}, \citenamefont {Donev},\ and\ \citenamefont {Garcia}}]{Pera6}%
  \BibitemOpen
  \bibfield  {author} {\bibinfo {author} {\bibfnamefont {J.-P.}\ \bibnamefont
  {P{\'e}raud}}, \bibinfo {author} {\bibfnamefont {A.}~\bibnamefont {Nonaka}},
  \bibinfo {author} {\bibfnamefont {A.}~\bibnamefont {Chaudhri}}, \bibinfo
  {author} {\bibfnamefont {J.~B.}\ \bibnamefont {Bell}}, \bibinfo {author}
  {\bibfnamefont {A.}~\bibnamefont {Donev}}, \ and\ \bibinfo {author}
  {\bibfnamefont {A.~L.}\ \bibnamefont {Garcia}},\ }\bibfield  {title}
  {\enquote {\bibinfo {title} {Low {M}ach number fluctuating hydrodynamics for
  electrolytes},}\ }\href@noop {} {\bibfield  {journal} {\bibinfo  {journal}
  {Physical Review Fluids}\ }\textbf {\bibinfo {volume} {1}},\ \bibinfo {pages}
  {074103} (\bibinfo {year} {2016})}\BibitemShut {NoStop}%
\bibitem [{\citenamefont {P{\'e}raud}\ \emph {et~al.}(2017)\citenamefont
  {P{\'e}raud}, \citenamefont {Nonaka}, \citenamefont {Bell}, \citenamefont
  {Donev},\ and\ \citenamefont {Garcia}}]{Pera5}%
  \BibitemOpen
  \bibfield  {author} {\bibinfo {author} {\bibfnamefont {J.-P.}\ \bibnamefont
  {P{\'e}raud}}, \bibinfo {author} {\bibfnamefont {A.~J.}\ \bibnamefont
  {Nonaka}}, \bibinfo {author} {\bibfnamefont {J.~B.}\ \bibnamefont {Bell}},
  \bibinfo {author} {\bibfnamefont {A.}~\bibnamefont {Donev}}, \ and\ \bibinfo
  {author} {\bibfnamefont {A.~L.}\ \bibnamefont {Garcia}},\ }\bibfield  {title}
  {\enquote {\bibinfo {title} {Fluctuation-enhanced electric conductivity in
  electrolyte solutions},}\ }\href@noop {} {\bibfield  {journal} {\bibinfo
  {journal} {Proceedings of the National Academy of Sciences}\ ,\ \bibinfo
  {pages} {201714464}} (\bibinfo {year} {2017})}\BibitemShut {NoStop}%
\bibitem [{\citenamefont {Williams}, \citenamefont {Bell},\ and\ \citenamefont
  {Garcia}(2008)}]{Will1}%
  \BibitemOpen
  \bibfield  {author} {\bibinfo {author} {\bibfnamefont {S.~A.}\ \bibnamefont
  {Williams}}, \bibinfo {author} {\bibfnamefont {J.~B.}\ \bibnamefont {Bell}},
  \ and\ \bibinfo {author} {\bibfnamefont {A.~L.}\ \bibnamefont {Garcia}},\
  }\bibfield  {title} {\enquote {\bibinfo {title} {Algorithm refinement for
  fluctuating hydrodynamics},}\ }\href@noop {} {\bibfield  {journal} {\bibinfo
  {journal} {Multiscale Modeling \& Simulation}\ }\textbf {\bibinfo {volume}
  {6}},\ \bibinfo {pages} {1256--1280} (\bibinfo {year} {2008})}\BibitemShut
  {NoStop}%
\bibitem [{\citenamefont {Donev}\ \emph
  {et~al.}(2010{\natexlab{b}})\citenamefont {Donev}, \citenamefont {Bell},
  \citenamefont {Garcia},\ and\ \citenamefont {Alder}}]{Done2}%
  \BibitemOpen
  \bibfield  {author} {\bibinfo {author} {\bibfnamefont {A.}~\bibnamefont
  {Donev}}, \bibinfo {author} {\bibfnamefont {J.~B.}\ \bibnamefont {Bell}},
  \bibinfo {author} {\bibfnamefont {A.~L.}\ \bibnamefont {Garcia}}, \ and\
  \bibinfo {author} {\bibfnamefont {B.~J.}\ \bibnamefont {Alder}},\ }\bibfield
  {title} {\enquote {\bibinfo {title} {A hybrid particle-continuum method for
  hydrodynamics of complex fluids},}\ }\href@noop {} {\bibfield  {journal}
  {\bibinfo  {journal} {Multiscale Modeling \& Simulation}\ }\textbf {\bibinfo
  {volume} {8}},\ \bibinfo {pages} {871--911} (\bibinfo {year}
  {2010}{\natexlab{b}})}\BibitemShut {NoStop}%
\bibitem [{\citenamefont {Cleuren}, \citenamefont {Van~den Broeck},\ and\
  \citenamefont {Kawai}(2006)}]{Broe1}%
  \BibitemOpen
  \bibfield  {author} {\bibinfo {author} {\bibfnamefont {B.}~\bibnamefont
  {Cleuren}}, \bibinfo {author} {\bibfnamefont {C.}~\bibnamefont {Van~den
  Broeck}}, \ and\ \bibinfo {author} {\bibfnamefont {R.}~\bibnamefont
  {Kawai}},\ }\bibfield  {title} {\enquote {\bibinfo {title} {Fluctuation
  theorem for the effusion of an ideal gas},}\ }\href {\doibase
  10.1103/PhysRevE.74.021117} {\bibfield  {journal} {\bibinfo  {journal} {Phys.
  Rev. E}\ }\textbf {\bibinfo {volume} {74}},\ \bibinfo {pages} {021117}
  (\bibinfo {year} {2006})}\BibitemShut {NoStop}%
\bibitem [{\citenamefont {Alexander}\ and\ \citenamefont
  {Garcia}(1997)}]{Alex2}%
  \BibitemOpen
  \bibfield  {author} {\bibinfo {author} {\bibfnamefont {F.~J.}\ \bibnamefont
  {Alexander}}\ and\ \bibinfo {author} {\bibfnamefont {A.~L.}\ \bibnamefont
  {Garcia}},\ }\bibfield  {title} {\enquote {\bibinfo {title} {The direct
  simulation {M}onte {C}arlo method},}\ }\href@noop {} {\bibfield  {journal}
  {\bibinfo  {journal} {Computers in Physics}\ }\textbf {\bibinfo {volume}
  {11}},\ \bibinfo {pages} {588--593} (\bibinfo {year} {1997})}\BibitemShut
  {NoStop}%
\bibitem [{\citenamefont {Garcia}(2000)}]{Garc2}%
  \BibitemOpen
  \bibfield  {author} {\bibinfo {author} {\bibfnamefont {A.~L.}\ \bibnamefont
  {Garcia}},\ }\href@noop {} {\emph {\bibinfo {title} {Numerical methods for
  physics}}}\ (\bibinfo  {publisher} {Prentice Hall Englewood Cliffs, NJ},\
  \bibinfo {year} {2000})\BibitemShut {NoStop}%
\bibitem [{\citenamefont {Boyd}\ and\ \citenamefont
  {Schwartzentruber}(2017)}]{Boyd1}%
  \BibitemOpen
  \bibfield  {author} {\bibinfo {author} {\bibfnamefont {I.}~\bibnamefont
  {Boyd}}\ and\ \bibinfo {author} {\bibfnamefont {T.}~\bibnamefont
  {Schwartzentruber}},\ }\href {https://books.google.com/books?id=tllEDgAAQBAJ}
  {\emph {\bibinfo {title} {Nonequilibrium Gas Dynamics and Molecular
  Simulation}}},\ Cambridge Aerospace Series\ (\bibinfo  {publisher} {Cambridge
  University Press},\ \bibinfo {year} {2017})\BibitemShut {NoStop}%
\bibitem [{\citenamefont {Keizer}(1978)}]{Keiz1}%
  \BibitemOpen
  \bibfield  {author} {\bibinfo {author} {\bibfnamefont {J.}~\bibnamefont
  {Keizer}},\ }\bibfield  {title} {\enquote {\bibinfo {title} {A theory of
  spontaneous fluctuations in viscous fluids far from equilibrium},}\ }\href
  {\doibase 10.1063/1.862214} {\bibfield  {journal} {\bibinfo  {journal} {The
  Physics of Fluids}\ }\textbf {\bibinfo {volume} {21}},\ \bibinfo {pages}
  {198--208} (\bibinfo {year} {1978})}\BibitemShut {NoStop}%
\bibitem [{\citenamefont {Keizer}(2012)}]{Keiz2}%
  \BibitemOpen
  \bibfield  {author} {\bibinfo {author} {\bibfnamefont {J.}~\bibnamefont
  {Keizer}},\ }\href {https://books.google.com/books?id=Gx8LCAAAQBAJ} {\emph
  {\bibinfo {title} {Statistical Thermodynamics of Nonequilibrium Processes}}}\
  (\bibinfo  {publisher} {Springer New York},\ \bibinfo {year}
  {2012})\BibitemShut {NoStop}%
\bibitem [{\citenamefont {Hadjiconstantinou}\ \emph {et~al.}(2003)\citenamefont
  {Hadjiconstantinou}, \citenamefont {Garcia}, \citenamefont {Bazant},\ and\
  \citenamefont {He}}]{Hadj6}%
  \BibitemOpen
  \bibfield  {author} {\bibinfo {author} {\bibfnamefont {N.~G.}\ \bibnamefont
  {Hadjiconstantinou}}, \bibinfo {author} {\bibfnamefont {A.~L.}\ \bibnamefont
  {Garcia}}, \bibinfo {author} {\bibfnamefont {M.~Z.}\ \bibnamefont {Bazant}},
  \ and\ \bibinfo {author} {\bibfnamefont {G.}~\bibnamefont {He}},\ }\bibfield
  {title} {\enquote {\bibinfo {title} {Statistical error in particle
  simulations of hydrodynamic phenomena},}\ }\href@noop {} {\bibfield
  {journal} {\bibinfo  {journal} {Journal of Computational Physics}\ }\textbf
  {\bibinfo {volume} {187}},\ \bibinfo {pages} {274--297} (\bibinfo {year}
  {2003})}\BibitemShut {NoStop}%
\bibitem [{\citenamefont {Mansour}\ \emph {et~al.}(1987)\citenamefont
  {Mansour}, \citenamefont {Garcia}, \citenamefont {Lie},\ and\ \citenamefont
  {Clementi}}]{Mans1}%
  \BibitemOpen
  \bibfield  {author} {\bibinfo {author} {\bibfnamefont {M.~M.}\ \bibnamefont
  {Mansour}}, \bibinfo {author} {\bibfnamefont {A.~L.}\ \bibnamefont {Garcia}},
  \bibinfo {author} {\bibfnamefont {G.~C.}\ \bibnamefont {Lie}}, \ and\
  \bibinfo {author} {\bibfnamefont {E.}~\bibnamefont {Clementi}},\ }\bibfield
  {title} {\enquote {\bibinfo {title} {Fluctuating hydrodynamics in a dilute
  gas},}\ }\href {\doibase 10.1103/PhysRevLett.58.874} {\bibfield  {journal}
  {\bibinfo  {journal} {Phys. Rev. Lett.}\ }\textbf {\bibinfo {volume} {58}},\
  \bibinfo {pages} {874--877} (\bibinfo {year} {1987})}\BibitemShut {NoStop}%
\bibitem [{\citenamefont {Garcia}\ and\ \citenamefont {Penland}(1991)}]{Garc4}%
  \BibitemOpen
  \bibfield  {author} {\bibinfo {author} {\bibfnamefont {A.}~\bibnamefont
  {Garcia}}\ and\ \bibinfo {author} {\bibfnamefont {C.}~\bibnamefont
  {Penland}},\ }\bibfield  {title} {\enquote {\bibinfo {title} {Fluctuating
  hydrodynamics and principal oscillation pattern analysis},}\ }\href@noop {}
  {\bibfield  {journal} {\bibinfo  {journal} {Journal of Statistical Physics}\
  }\textbf {\bibinfo {volume} {64}},\ \bibinfo {pages} {1121--1132} (\bibinfo
  {year} {1991})}\BibitemShut {NoStop}%
\bibitem [{\citenamefont {Bruno}\ \emph {et~al.}(2006)\citenamefont {Bruno},
  \citenamefont {Capitelli}, \citenamefont {Longo},\ and\ \citenamefont
  {Minelli}}]{Bruno1}%
  \BibitemOpen
  \bibfield  {author} {\bibinfo {author} {\bibfnamefont {D.}~\bibnamefont
  {Bruno}}, \bibinfo {author} {\bibfnamefont {M.}~\bibnamefont {Capitelli}},
  \bibinfo {author} {\bibfnamefont {S.}~\bibnamefont {Longo}}, \ and\ \bibinfo
  {author} {\bibfnamefont {P.}~\bibnamefont {Minelli}},\ }\bibfield  {title}
  {\enquote {\bibinfo {title} {Monte carlo simulation of light scattering
  spectra in atomic gases},}\ }\href@noop {} {\bibfield  {journal} {\bibinfo
  {journal} {Chemical Physics Letters}\ }\textbf {\bibinfo {volume} {422}},\
  \bibinfo {pages} {571 -- 574} (\bibinfo {year} {2006})}\BibitemShut {NoStop}%
\bibitem [{\citenamefont {Bruno}, \citenamefont {Frezzotti},\ and\
  \citenamefont {Ghiroldi}(2017)}]{Bruno2}%
  \BibitemOpen
  \bibfield  {author} {\bibinfo {author} {\bibfnamefont {D.}~\bibnamefont
  {Bruno}}, \bibinfo {author} {\bibfnamefont {A.}~\bibnamefont {Frezzotti}}, \
  and\ \bibinfo {author} {\bibfnamefont {G.~P.}\ \bibnamefont {Ghiroldi}},\
  }\bibfield  {title} {\enquote {\bibinfo {title} {Rayleigh–brillouin
  scattering in molecular oxygen by ct-dsmc simulations},}\ }\href@noop {}
  {\bibfield  {journal} {\bibinfo  {journal} {European Journal of Mechanics -
  B/Fluids}\ }\textbf {\bibinfo {volume} {64}},\ \bibinfo {pages} {8 -- 16}
  (\bibinfo {year} {2017})}\BibitemShut {NoStop}%
\bibitem [{\citenamefont {Giovangigli}(2012)}]{Giov1}%
  \BibitemOpen
  \bibfield  {author} {\bibinfo {author} {\bibfnamefont {V.}~\bibnamefont
  {Giovangigli}},\ }\href {https://books.google.com/books?id=wesHCAAAQBAJ}
  {\emph {\bibinfo {title} {Multicomponent Flow Modeling}}},\ Modeling and
  Simulation in Science, Engineering and Technology\ (\bibinfo  {publisher}
  {Birkh{\"a}user Boston},\ \bibinfo {year} {2012})\BibitemShut {NoStop}%
\bibitem [{\citenamefont {Liepmann}(1961)}]{Liep1}%
  \BibitemOpen
  \bibfield  {author} {\bibinfo {author} {\bibfnamefont {H.~W.}\ \bibnamefont
  {Liepmann}},\ }\bibfield  {title} {\enquote {\bibinfo {title} {Gaskinetics
  and gasdynamics of orifice flow},}\ }\href {\doibase
  10.1017/S002211206100007X} {\bibfield  {journal} {\bibinfo  {journal}
  {Journal of Fluid Mechanics}\ }\textbf {\bibinfo {volume} {10}},\ \bibinfo
  {pages} {65–79} (\bibinfo {year} {1961})}\BibitemShut {NoStop}%
\bibitem [{\citenamefont {Gentle}(2013)}]{Gent1}%
  \BibitemOpen
  \bibfield  {author} {\bibinfo {author} {\bibfnamefont {J.}~\bibnamefont
  {Gentle}},\ }\href {https://books.google.com/books?id=KJzlBwAAQBAJ} {\emph
  {\bibinfo {title} {Random Number Generation and Monte Carlo Methods}}},\
  Statistics and Computing\ (\bibinfo  {publisher} {Springer New York},\
  \bibinfo {year} {2013})\BibitemShut {NoStop}%
\bibitem [{\citenamefont {Gillespie}(2007)}]{Gill1}%
  \BibitemOpen
  \bibfield  {author} {\bibinfo {author} {\bibfnamefont {D.~T.}\ \bibnamefont
  {Gillespie}},\ }\bibfield  {title} {\enquote {\bibinfo {title} {Stochastic
  simulation of chemical kinetics},}\ }\href@noop {} {\bibfield  {journal}
  {\bibinfo  {journal} {Annual Review of Physical Chemistry}\ }\textbf
  {\bibinfo {volume} {58}},\ \bibinfo {pages} {35--55} (\bibinfo {year}
  {2007})}\BibitemShut {NoStop}%
\bibitem [{\citenamefont {Garcia}(2007)}]{Garc3}%
  \BibitemOpen
  \bibfield  {author} {\bibinfo {author} {\bibfnamefont {A.}~\bibnamefont
  {Garcia}},\ }\bibfield  {title} {\enquote {\bibinfo {title} {Estimating
  hydrodynamic quantities in the presence of microscopic fluctuations},}\
  }\href@noop {} {\bibfield  {journal} {\bibinfo  {journal} {Communications in
  Applied Mathematics and Computational Science}\ }\textbf {\bibinfo {volume}
  {1}},\ \bibinfo {pages} {53--78} (\bibinfo {year} {2007})}\BibitemShut
  {NoStop}%
\bibitem [{\citenamefont {Garcia}\ \emph {et~al.}(1987)\citenamefont {Garcia},
  \citenamefont {Mansour}, \citenamefont {Lie},\ and\ \citenamefont
  {Cementi}}]{Garcia1987}%
  \BibitemOpen
  \bibfield  {author} {\bibinfo {author} {\bibfnamefont {A.~L.}\ \bibnamefont
  {Garcia}}, \bibinfo {author} {\bibfnamefont {M.~M.}\ \bibnamefont {Mansour}},
  \bibinfo {author} {\bibfnamefont {G.~C.}\ \bibnamefont {Lie}}, \ and\
  \bibinfo {author} {\bibfnamefont {E.}~\bibnamefont {Cementi}},\ }\bibfield
  {title} {\enquote {\bibinfo {title} {Numerical integration of the fluctuating
  hydrodynamic equations},}\ }\href@noop {} {\bibfield  {journal} {\bibinfo
  {journal} {Journal of Statistical Physics}\ }\textbf {\bibinfo {volume}
  {47}},\ \bibinfo {pages} {209--228} (\bibinfo {year} {1987})}\BibitemShut
  {NoStop}%
\bibitem [{\citenamefont {Nicolis}\ and\ \citenamefont
  {Mansour}(1984)}]{Nico1}%
  \BibitemOpen
  \bibfield  {author} {\bibinfo {author} {\bibfnamefont {G.}~\bibnamefont
  {Nicolis}}\ and\ \bibinfo {author} {\bibfnamefont {M.~M.}\ \bibnamefont
  {Mansour}},\ }\bibfield  {title} {\enquote {\bibinfo {title} {Onset of
  spatial correlations in nonequilibrium systems: a master-equation
  description},}\ }\href@noop {} {\bibfield  {journal} {\bibinfo  {journal}
  {Physical Review A}\ }\textbf {\bibinfo {volume} {29}},\ \bibinfo {pages}
  {2845} (\bibinfo {year} {1984})}\BibitemShut {NoStop}%
\end{thebibliography}%

\end{document}